\documentclass[journal]{new-aiaa}
\usepackage[utf8]{inputenc}
\usepackage{textcomp}

\usepackage{graphicx}
\usepackage{amsmath}
\usepackage[version=4]{mhchem}
\usepackage{siunitx}

\usepackage{longtable,tabularx}
\graphicspath{{./}}

\usepackage{hyperref}
\hypersetup{
    hidelinks
}

\usepackage{url}
\usepackage{subcaption}
\usepackage{enumitem}

\usepackage{amsmath}

\usepackage{bm}

\usepackage{multirow}
\usepackage{ulem}

\usepackage{lipsum} % Use, e.g.: \lipsum[3-5]
\usepackage{xcolor} % Use, e.g.: \textcolor{red}{...}
\usepackage{todonotes}
\usepackage{lineno} 

\usepackage[version=4]{mhchem}
\usepackage{siunitx}
\usepackage{longtable,tabularx}
\title{Learning neural-network-based turbulence models for external transonic flows using ensemble Kalman method}

\author{Yi Liu \footnote{Postdoctoral Research Associate, LNM, Institute of Mechanics; liuyi@imech.ac.cn }}

\author{Xin-Lei Zhang \footnote{Postdoctoral Research Associate, LNM, Institute of Mechanics; zhangxinlei@imech.ac.cn (Corresponding author).}}

\author{Guowei He \footnote{Professor, LNM, Institute of Mechanics; hgw@lnm.imech.ac.cn (Corresponding author).}}

\affil{State Key Laboratory of Nonlinear Mechanics, Chinese Academy of Sciences, 100190 Beijing, \\ People’s Republic of China}

\affil{University of Chinese Academy of Sciences, 100049 Beijing, People’s Republic of China}

\begin{document}

\maketitle

    \begin{abstract}
        This paper presents a neural network-based turbulence modeling approach for transonic flows based on the ensemble Kalman method. The approach adopts a tensor basis neural network for the Reynolds stress representation, with modified inputs to consider fluid compressibility.
        The normalization of input features is also investigated to avoid feature collapsing in the presence of shock waves. Moreover, the turbulent heat flux is accordingly estimated with the neural network-based turbulence model based on the gradient diffusion hypothesis. The ensemble Kalman method is used to train the neural network with the experimental data in velocity and wall pressure due to its derivative-free nature. The proposed framework is tested in two canonical configurations, i.e., 2D transonic flows over the RAE2822 airfoils and 3D transonic flows over the ONERA M6 wings. Numerical results demonstrate the capability of the proposed method in learning accurate turbulence models for external transonic flows.
    \end{abstract}

\section*{Nomenclature}

{
\renewcommand\arraystretch{1.0}
\noindent\begin{longtable*}{@{}l @{\quad=\quad} l@{}}
$a$     &  speed of sound   \\
$C_f$   & friction coefficient \\
$C_p$   & pressure coefficient \\
$c_p$   & specific heat coefficient at constant pressure   \\
$C$     & chord length \\
$E$     &  total energy   \\
${g}$  &  coefficient of tensor basis \\
$H$          &  total enthalpy  \\
${\mathbf{I}}$  &  identity matrix  \\
$k$             &  turbulent kinetic energy \\
$L$             &  reference length \\
$M$             &  number of samples \\
$Ma$            &  Mach number    \\
$Ma_n$          &  normal Mach number    \\
$p$             &  static pressure \\
$\mathsf{P}$    &  model error covariance   \\
$Pr$            &  Prandtl number  \\
${q_j},q_j^{(t)}$  &  laminar heat flux and turbulent heat flux  \\
$\mathsf{R}$       &  observation error covariance  \\
$Re$  & Reynolds number \\   
${\mathbf{S}},{\Omega}$  & strain-rate tensor and rotation-rate tensor \\  
$\hat{\mathbf{S}}, \hat{\Omega}$  & non-dimensionalized strain-rate tensor and rotation-rate tensor \\ 
$T$    &  static temperature  \\
${T_\infty }$ & temperature at far-field \\  
${{\mathbf{T}}}$ &  tensor basis \\
${U_\infty }$   &  free-stream velocity  \\ 
${\mathbf{u}}$  &  velocity vector \\
% $u,v,w$       &  velocity components in the $x$-, $y$-, $z$- direction \\
$w$       &  weights of neural networks \\
$W$       &  ensemble of weights, $W=\{ w \}_{m=1}^M$ \\
% $u_i$           &  velocity components $(u,v,w)$ in Cartesian directions $x,y,z$ \\
% $u_b$           &  bulk velocity  \\
$x$   &  Cartesian coordinates \\
$\mathsf{y}$ & training data \\
$y^+$   &  nondimensional wall distance \\
$\alpha$   &  angle of attack  \\
$\beta ^*$  & turbulence model constants  \\
$\mathcal{H}$     &  model operator that maps the neural network weights to the observed quantities \\
${\mu, \mu _t }$  &  dynamic viscosity and turbulent viscosity  \\
${\theta}$     &  scalar invariants \\
$\rho$            &  fluid density  \\
$\sigma$     &  viscous stress \\
$\tau$            &  Reynolds stress  \\
$\omega$      &  specific dissipation rate of turbulent kinetic energy  \\
${\nabla}$  &  gradient operator \\
$\|*\|$       &  L2 norm  \\
$\text{tr}\left\{ * \right\}$  & trace operator  \\

\multicolumn{2}{@{}l}{Superscripts}\\
${(i)}$ & index of tensor basis \\
$n$ & index of learning iteration \\
${\top}$ & transpose \\
${0}$ & prior \\

\multicolumn{2}{@{}l}{Subscripts}\\
${i,j,k}$  & 1, 2, 3, tensor indices \\ %direction in computational space \\
${m}$ & index of sample \\
${n}$ & normal component of vector  \\
${t}$      & turbulent \\
$MAC$      & mean aerodynamic chord \\
$RSM$      & Reynolds-stress transport model  \\
${\infty}$ & at infinity \\

\end{longtable*}}

\section{Introduction}
Accurate prediction of transonic flows is of significant interest for prior assessments of buffet onset, which further concerns the safety design of flight vehicles~\cite{jameson201150,giannelis2017review}.
The Reynolds averaged Navier--Stokes (RANS) simulation with turbulence models is still the workhorse tool for predicting such flows with high Reynolds numbers due to its low computational overhead~\cite{pi2014cfd}.
Therefore, it is necessary to develop accurate turbulence models for transonic flows.

The most widely used turbulence models are linear eddy viscosity models, such as Spalart--Allmaras model~\cite{spalart1992one-equation} and $k$--$\omega$ SST model~\cite{menter1994two}.
These models assume that the Reynolds stress is linearly related to the strain rate via turbulent eddy viscosity based on the Boussinesq assumption.
They can provide satisfactory predictions for attached flows but often lead to large discrepancies when encountered with the mean curvature and separation~\cite{craft1996development,mompean1996predicting}.
For transonic flows, there often occur turbulent heat transfer and shock waves, which would lead to additional difficulties in turbulence modeling.
Specifically, the turbulent heat transfer requires closure modeling for turbulent heat flux, which is usually estimated based on the gradient diffusion hypothesis. 
That is, the turbulent heat flux is proportional to the temperature gradient, while the proportional coefficient involves the turbulent eddy viscosity obtained with turbulence models.
Hence the turbulence model has to be used for estimating the Reynolds stress as well as turbulent heat flux.
Moreover, the transonic flows would produce shock waves, which often interact with the boundary layer and lead to flow separations.
Estimating such shock-induced separation requires capturing the interaction between the shock wave and boundary layer, which poses another challenge for turbulence models.
Some advanced turbulence models that go beyond the Boussinesq assumption, e.g., the Reynolds stress transport model, can provide satisfactory predictions for transonic flows but encounter the robustness issue and increased computational cost~\cite{wilcox1998turbulence}.
The compromise between predictive accuracy and practical usability is often intractable for turbulence modeling.

Machine learning methods~\cite{brunton2020machine,sun2020physics,WuICA} have been increasingly used to construct high-fidelity turbulence models due to their ability of discovering complicated functional relationships from data.
For instance, the tensor basis neural network~\cite{ling2016reynolds} and the symbolic expression~\cite{weatheritt2016novel} are proposed to represent the Reynolds stress in an algebraic form based on the weak-equilibrium assumption.
Also, the vector-cloud neural network~\cite{zhou2021learning} is proposed to emulate the Reynolds stress transport equations with embedded non-equilibrium effects.
However, these methods usually require full field data of Reynolds stresses to determine the model parameters, i.e., the weights of neural networks.
These Reynolds stress data can be categorized as direct data for learning turbulence models.
This is in contrast to the indirect data~\cite{strofer2021end} such as velocity and wall pressure, which need to be propagated from the Reynolds stress via the RANS equation.
Learning from the direct data would cause poor posterior prediction due to the inconsistency between training and prediction environments~\cite{duraisamy2021perspectives}.
In addition, the full field data of Reynolds stress is difficult to be obtained for transonic flows at high Reynolds numbers.
In view of these difficulties, learning turbulence models from indirect data becomes trending since it can reduce the data requirement and avoid the inconsistency issue simultaneously.
Such a strategy is often referred to as model-consistent training~\cite{duraisamy2021perspectives}.

Model-consistent training ensures consistency between the training and prediction environments by involving RANS equations in the training process.
Different strategies, such as adjoint-based learning~\cite{singh2017machine,holland2019field,strofer2021end}, ensemble-based learning~\cite{strofer2021dafi, zhang2022ensemble}, and genetic programming~\cite{zhao2020rans}, have been proposed but are mostly applied for low-speed incompressible flows.
Over the past few years, the genetic programming method has been increasingly used to learn turbulence models for transonic flows in a model-consistent manner due to its gradient-free merits. 
That is, the gradient evaluation is not needed to minimize the discrepancies between the training data and model predictions as the adjoint-based method. 
This method has been used to construct  algebraic Reynolds stress models in symbolic forms  for the wake of turbine cascade~\cite{zhao2020rans} and the separation-induced transition~\cite{akolekar2021transition} in turbomachines.
In contrast to the gradient-free method, the ensemble Kalman method is a gradient-approximation method, which estimates the model gradient and Hessian based on the statistics of random samples.
By using the approximated gradient and Hessian information, such methods are able to achieve good training efficiency and also do not require developing adjoint solvers.
The ensemble Kalman method is recently used to learn neural network-based turbulence models from indirect data~\cite{zhang2022ensemble} and shows superior efficiency and considerable training accuracy compared to the adjoint-based learning method.
The ensemble-based method has only been demonstrated for incompressible flows, and its application for transonic flows is worthy of further investigation.

The present work aims to develop the neural network-based turbulence modeling framework for transonic flows based on the ensemble Kalman method.
Most neural network-based turbulence models are constructed for low-speed flows.
Transonic flows have different flow features from low-speed flows due to fluid compressibility, shock waves, and heat transfer, which would require specialized treatments for the neural network-based representation of Reynolds stresses.
In this work, the compressibility effects are taken into account in selecting the input features of neural networks.
Further, the scaling strategy of input features is investigated to avoid  feature collapse due to the presence of shock waves.
Moreover, the heat flux closure is modified with the outputs of neural networks based on the Boussinesq hypothesis.
The ensemble Kalman method is used to train the neural network-based model with good training efficiency and flexibility in data usage.
Finally, the method is tested for transonic flows over the ONERA M6 wing, demonstrating its applicability in 3D flow applications.

The rest of the paper is outlined as follows.
The tensor-basis neural network for transonic flows and the ensemble-based training algorithm are introduced in Section~\ref{sec:2-NumericalMethod}.
The test cases and corresponding results are presented and analyzed in Section~\ref{sec:3-NumericalResults}.
Finally, this paper is concluded in Section~\ref{sec:4-Conclusion}.

\section{Ensemble-based turbulence modeling for transonic flows \label{sec:2-NumericalMethod} }
The Reynolds-averaged Navier-Stokes (RANS) equations for compressible flows can be written as~\cite{wilcox1998turbulence}
    \begin{subequations}
    \label{eq:NSEquations}
    \begin{equation}
        \frac{{\partial \rho }}{{\partial t}} + \frac{{\partial (\rho {u_j})}}{{\partial {x_j}}} = 0 ,\label{subeq:1}
    \end{equation}
    \begin{equation}
        \frac{{\partial (\rho {u_i})}}{{\partial t}} + \frac{\partial }{{\partial {x_j}}}(\rho {u_i}{u_j}) =  - \frac{{\partial p}}{{\partial {x_i}}} + \frac{{\partial {\sigma _{ij}}}}{{\partial {x_j}}} + \frac{{\partial {\tau _{ij}}}}{{\partial {x_j}}},\label{subeq:2}
    \end{equation}
    \begin{equation}
        \frac{{\partial (\rho E)}}{{\partial t}} + \frac{\partial }{{\partial {x_j}}}(\rho H{u_j}) = \frac{{\partial ({\tau _{ij}}{u_i})}}{{\partial {x_j}}} + \frac{{\partial ({\sigma _{ij}}{u_i})}}{{\partial {x_j}}} - \frac{{\partial {q_j}}}{{\partial {x_k}}} - \frac{{\partial q_j^{\left( t \right)}}}{{\partial {x_j}}}. \label{subeq:3}
    \end{equation}
    \end{subequations}
In the formula above, $\rho$ and $p$ represent the density and pressure, respectively, $u_i$ is the flow velocity component, $\tau_{ij}$ is the Reynolds stress tensor, $\sigma_{ij}$ represents the viscous stress, which is described by 
$${\sigma _{ij}} = \mu \left( {\frac{{\partial {u_i}}}{{\partial {x_j}}} + \frac{{\partial {u_j}}}{{\partial {x_i}}} - \frac{2}{3}\frac{{\partial {u_k}}}{{\partial {x_k}}}{\delta _{ij}}} \right), $$
and $\mu$ is the dynamic viscosity.
In the energy equation~\eqref{subeq:3}, $E$ and $H$ denote the total energy and total enthalpy, ${q_j^{\left(t \right)}}$ is the turbulent heat flux,
and ${q_j}$ is the laminar heat flux.
The heat flux ${q_j}$ is calculated by $$ {q_j} =  - \frac{{{c_p}\mu }}{{Pr }}\frac{{\partial T}}{{\partial {x_j}}} ,$$ 
where $T$ is the temperature, and ${c_p}$ and ${Pr}$ are specific heat coefficients at constant pressure and Prandtl number, respectively.
To close these equations, the Reynolds stress~$\tau_{ij}$ and turbulent heat flux~${q_j^{\left(t \right)}}$ need to be modeled, which will be introduced in the following.

\subsection{Reynolds stress closure with tensor basis neural network}
    We consider the nonlinear eddy viscosity model as the baseline model to represent the Reynolds stresses~\cite{ling2016reynolds}.
    Such a choice is based on the following considerations.
    First, the nonlinear eddy viscosity model uses the integrity of tensor bases which is more flexible in contrast to the linear eddy viscosity model that uses only the first tensor basis.
    Moreover, the model is straightforward to be implemented and computationally efficient since no additional transport equations need to be solved.
    Besides, the nonlinear eddy viscosity models have often been used for numerical predictions of transonic flows~\cite{barakos2000numerical,barakos2000investigation}.

    The nonlinear eddy viscosity model represents the deviatoric part of Reynolds stress with ten independent tensor bases based on Cayley-Hamilton theory~\cite{pope2000turbulent}.
    Specifically, the Reynolds stress can be formulated as
    \begin{equation}
        {\mathbf{\tau}} = -2\rho k\sum\limits_{i = 1}^{10} {{g^{(i)}}} {{\mathbf{T}}^{(i)}} - \frac{{2\rho k}}{3}{\mathbf{I}} \text{,}
    \label{eq:ReStress}
    \end{equation}
    where $k$ is turbulent kinetic energy, ${\mathbf{I}}$ is identity matrix, and
    ${g^{(i)}}$ is coefficient of the tensor basis ${{\mathbf{T}}^{(i)}}$.
    The coefficient $\mathbf{g}$ can be represented as the function of scalar invariants~${\mathbf{\theta}}$ of these tensor bases.
    For compressible flows, we have to consider the non-zero trace of the strain rate, and the tensor bases can be written as~\cite{wallin2000explicit,abdol2018development}
    \begin{equation}
    \label{eq:tensor}
    \centering
    \begin{gathered}
    \begin{array}{*{20}{c}}
        {{{\mathbf{T}}^{\left( 1 \right)}} = \left[ {\hat {\mathbf{S}} - \frac{1}{3}{\text{tr}}\left\{ {\hat {\mathbf{S}}} \right\}} \right]}, & {{{\mathbf{T}}^{\left( 2 \right)}} = \left[ {\hat {\mathbf{S}}\hat \Omega  - \hat \Omega \hat {\mathbf{S}}} \right]}, & {{{\mathbf{T}}^{\left( 2 \right)}} = \left[ {{{{\mathbf{\hat S}}}^2} - \frac{1}{3}{\text{tr}}\left\{ {{{{\mathbf{\hat S}}}^2}} \right\}} \right]} \text{,}
    \end{array} \\ 
    \begin{array}{*{20}{c}}
         {{{\mathbf{T}}^{\left( 4 \right)}} = \left[ {{{\hat \Omega }^2} - \frac{1}{3}{\text{tr}}\left\{ {{{\hat \Omega }^2}} \right\}} \right]}, & {{{\mathbf{T}}^{\left( 5 \right)}} = \left[ {{{{\mathbf{\hat S}}}^2}\hat \Omega  - \hat \Omega {{{\mathbf{\hat S}}}^2}} \right]}, & {{{\mathbf{T}}^{\left( 6 \right)}} = \left[ {\hat {\mathbf{S}}{{\hat \Omega }^2} + {{\hat \Omega }^2}\hat {\mathbf{S}} - \frac{2}{3}{\text{tr}}\left\{ {\hat {\mathbf{S}}{{\hat \Omega }^2}} \right\}} \right]}   \text{,}
    \end{array} \\ 
    \begin{array}{*{20}{c}}
        {{{\mathbf{T}}^{\left( 7 \right)}} = \left[ {{{\hat {\mathbf{S}}}^2}{{\hat \Omega }^2} + {{\hat \Omega }^2}{{\hat {\mathbf{S}}}^2} - \frac{2}{3}{\text{tr}}\left\{ {{{{\mathbf{\hat S}}}^2}{{\hat \Omega }^2}} \right\}} \right]}, & {{{\mathbf{T}}^{\left( 8 \right)}} = \left[ {{\mathbf{\hat S}}\hat \Omega {{{\mathbf{\hat S}}}^2} - {{{\mathbf{\hat S}}}^2}\hat \Omega {\mathbf{\hat S}}} \right]}   \text{,}
    \end{array} \\ 
    \begin{array}{*{20}{c}}
        {{{\mathbf{T}}^{\left( 9 \right)}} = \left[ {\hat \Omega \hat {\mathbf{S}}{{\hat \Omega }^2} - {{\hat \Omega }^2}\hat {\mathbf{S}}\hat \Omega } \right]}, & {{{\mathbf{T}}^{\left( {10} \right)}} = \left[ {\hat \Omega {{{\mathbf{\hat S}}}^2}{{\hat \Omega }^2} - {{\hat \Omega }^2}{{{\mathbf{\hat S}}}^2}\hat \Omega } \right]}   \text{.}
    \end{array}
    \end{gathered}
\end{equation}
    Accordingly, one additional scalar invariant $\text{tr}{(\hat{\mathbf{S}})}$ need to be included compared to the incompressible scenario~\cite{ling2016reynolds}, 
    and hence there exist six scalar invariants as
\begin{equation}
    \label{eq:Inv}
    \centering
    \begin{gathered}
    \begin{array}{*{20}{c}}
      {{\theta _1} = \text{tr}\left\{ {\hat {\mathbf{S}}} \right\},} & {{\theta _2} = \text{tr}\left\{ {{{\hat {\mathbf{S}}}^2}} \right\},} & {{\theta _3} = \text{tr}\left\{ {{{\hat \Omega }^2}} \right\},} 
    \end{array} \\ 
    \begin{array}{*{20}{c}}
        {{\theta _4} = \text{tr}\left\{ {{{\hat {\mathbf{S}}}^3}} \right\},} & {{\theta _5} = \text{tr}\left\{ {\hat {\mathbf{S}}{{\hat \Omega }^2}} \right\},} & {{\theta _6} = \text{tr}\left\{ {{{\hat {\mathbf{S}}}^2}{{\hat \Omega }^2}} \right\}} 
    \end{array} { \text{.}}  
    \end{gathered}
\end{equation}
    In the formula above, $\hat{\mathbf{S}}$ and $\hat{\Omega}$ represent the non-dimensionalized strain rate and rotation rate
    with the turbulence time scale~$\tau_s$, i.e.,~$\hat{\mathbf{S}}=\tau_s \mathbf{S}$ and $\hat{\Omega}=\tau_s\Omega$.
    The turbulence time scale~$\tau_s$ can be estimated based on ${\tau_s = 1 / (\beta^* \omega})$, where ${\beta ^*}$ is a model constant, $\omega$ is specific turbulence dissipation rate that is obtained with $k$--$\omega$ SST model in this work.
    The formula of $k$--$\omega$ SST model is presented in Appendix~\ref{AppendixSST}.

    Conclusively, the coefficient $\mathbf{g}$ of tensor basis is represented with neural networks in this work, and the scalar invariants are used as the input features of the neural network.
    These inputs of the tensor basis neural network~\cite{ling2016reynolds} are modified for compressible flows to consider fluid compressibility.
    The obtained Reynolds stress tensor is used for the momentum equation, turbulent kinetic energy transport equation, and the energy equation in Eqs.~\eqref{eq:NSEquations}.

\subsection{Heat flux closure with tensor basis neural networks}
    The turbulent heat flux~$q^{(t)}$ requires to be modeled to close the energy equation~\eqref{subeq:3}.
    It is often estimated based on the gradient diffusion hypothesis. 
    That is, the heat is iso-tropically diffused as 
    \begin{equation}
    \label{eq:TurbHeatFlux}
    q_j^{\left( t \right)} =  - \alpha_t \frac{{\partial T}}{{\partial {x_j}}} \text{,}
    \end{equation}
where $\alpha_t$ is the turbulent thermal diffusivity. 
According to the Reynolds analogy between turbulent heat flux $q_j^{\left( t \right)}$ and momentum flux ${\mathbf{\tau}}$~\cite{geankoplis2003transport}, the diffusivity $\alpha_t$ can be calculated as 
    \begin{equation}
    \label{eq:Miut}
        \alpha_t =\frac{{{\mu _{t}}{c_p}}}{{P{r_{t}}}}
        \text{,} \quad \text{with} \quad
        {\mu _{{\text{t }}}} = -\frac{{{g^{\left( 1 \right)}}\rho k}}{{{\beta ^*}\omega }} \text{.}
    \end{equation}
Here $Pr_t$ is the turbulent Prandtl number, and $\mu_t$ is the turbulent eddy viscosity, which is estimated with the first term of the deviatoric part of Reynolds stress in Eq.~\eqref{eq:ReStress} by following the Bounssinesq assumption.
The eddy viscosity~$\mu_t$ is also used for the $k$ and $\omega$ transport equations.
Admittedly, such modification is still under the gradient diffusion hypothesis.
Future investigations need to represent heat flux closures based on the generalized gradient diffusion hypothesis~\cite{ling2016analysis}, which introduces the dependence on individual Reynolds stress components.
Moreover, the $g^{(1)}$ function is constructed with only scalar invariants of the velocity gradient. 
For the heat flux closure, the invariants of temperature gradient~\cite{milani2018machine,lav2021rans} should also be considered based on the representation theorems.
It is noted that introducing the invariants associated with temperature gradient would enable the heat flux model to be nonlinear.
Such nonlinear heat flux models are inappropriate for passive scalar transport equations in incompressible scenarios.
However, for compressible flows, the nonlinear heat flux could be applied since the temperature is not passive scalar due to coupling with momentum equations.
Also, we emphasize that this work mainly focuses on turbulence modeling, and the heat flux closure is modified according to the turbulence model.
Learning the heat flux closure from the heat flux data is of significant interest and will be pursued in the near future work.

Conclusively, we use the neural network to represent the mapping between the scalar invariant and the coefficient of tensor bases, i.e., $\boldsymbol{\theta} \mapsto \mathbf{g}$.
On the one hand, the obtained coefficients are combined with the tensor bases to form the Reynolds stress based on Eq.~\eqref{eq:ReStress}.
On the other hand, the first coefficient~$g^{(1)}$ is combined with the temperature gradient to estimate the turbulent heat flux based on Eq.~\eqref{eq:TurbHeatFlux}.
As such, the RANS equations are closed with the neural network-based closures.

\subsection{Normalization of the input features}
The scalar invariants~$\mathbf{\theta}$ are used as the input features of neural networks, which are often scaled in the range of, e.g., $[-1, 1]$, to improve the efficiency of the neural network training.
Two strategies, i.e., the min-max normalization and local normalization, are usually used to scale the input features.
The former approach uses the global minimum and maximum values to scale the input features, while the latter uses local quantities for scaling.
Both approaches have been employed for feature scaling of tensor basis neural networks~\cite{ling2015evaluation,strofer2021end}.

\begin{figure}[!htb]
    \centering
    \subfloat[RAE2822 airfoil]{
    \includegraphics[width=0.46\textwidth]{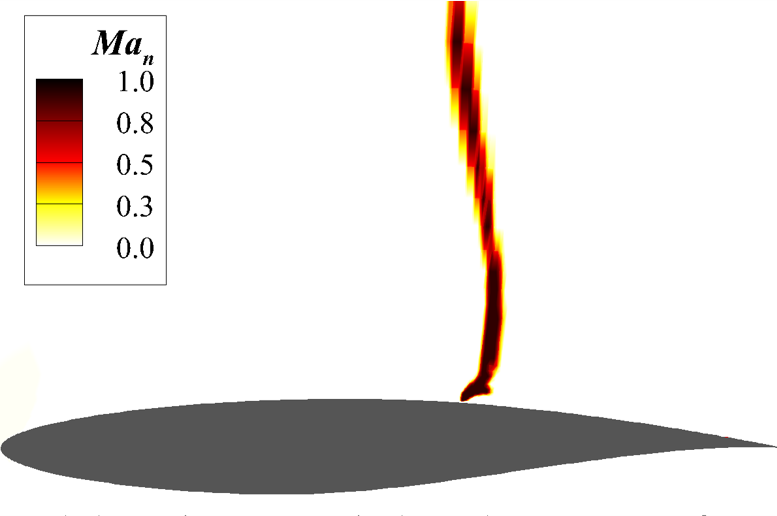} }
    \subfloat[ONERA M6 wing]{
    \includegraphics[width=0.46\textwidth]{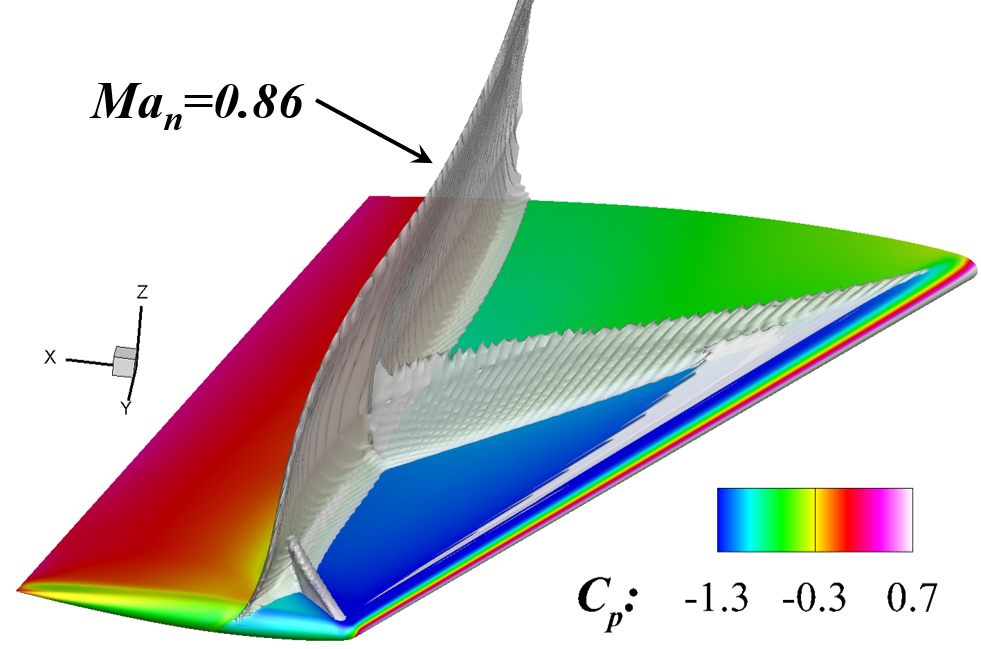} }
    \caption{Plots of the detected shock for RAE2822 case and ONERA M6 based on criteria of $Ma_n \geqslant 0.86$.
    }
    \label{fig:ShockDtec}
\end{figure}

The min-max normalization can ensure the input features within $[0, 1]$, which is widely used in the machine learning community.
However, when encountered with shock waves, local scalar invariants can be extremely large due to the sharp discontinuity in the velocity field, leading to feature collapses, as noted in ~\cite{zhang2022ensemble}.
To address this issue, the features in the shock wave regime can be identified and removed from the collection of input features.
Eliminating the features of shock waves has negligible effects on the model learning since the strong pressure gradients are dominant over the Reynolds stress in these areas.
One widely used criteria for stationary shock wave detection can be expressed as~\cite{lovely1999shock}
\begin{equation}
\begin{aligned}
    \label{eq:shockDetec}
    Ma_n > 0.86 \text{,} \quad \text{with} \quad Ma_n = \frac{{{\mathbf{u}} \cdot \nabla p}}{{a \left| {\nabla p} \right|}} \text{,}
\end{aligned}
\end{equation}
where $Ma_n$ is normal Mach number, ${\mathbf{u}}$ is the velocity vector of the local flow, and $a$ is the speed of sound.
The detected shock wave is shown in Fig.~\ref{fig:ShockDtec} for flows over the REA2822 airfoil and the ONERA M6 wings.
Based on the criteria, we can remove the features associated with the shock wave and thus eliminate their effects on the feature scaling.

Alternatively, the local normalization is also commonly used to scale the input features~$q$ based on $\hat{q} = q / (\| q \| + q^*)$~\cite{ling2015evaluation}, where $q^*$ is local quantities.
For the tensor-basis neural network, the time scale~$\tau_s$ is used as the local quantity for normalization.
Thus, we can normalize the strain rate and the rotation rate through
\begin{equation}
    \label{eq:nl}
    \hat{\mathbf{S}} = \frac{\mathbf{S}}{\|\mathbf{S}\|+1 / \tau_s} 
    \quad \text{and} \quad
    \hat{\mathbf{\Omega}} = \frac{\mathbf{\Omega}}{\|\mathbf{\Omega}\|+1 / \tau_s} \text{.}
\end{equation}
Further, the scalar invariants are computed with these normalized $\hat{\mathbf{S}}$ and $\hat{\mathbf{\Omega}}$ based on Eq.\eqref{eq:Inv}.
It is noted that this scaling is slightly different from the commonly used one, i.e.,~$\hat{q}=q/q^*$.
Because the choice of Eq.~\eqref{eq:nl} can ensure the input features within the range of~$[-1, 1]$, which can improve the convergence rate based on the practice of machine learning~\cite{ling2015evaluation, wu2018physics}.
Also, the range of input features is confined and known as a priori, which enables pretraining neural network-based models to be feasible.
The pertaining step is crucial to accelerate model learning and avoid nonphysical model outputs, such as negative eddy viscosity~\cite{zhang2022ensemble}, which will be illustrated in Section~\ref{sec:2-NumericalMethod}-D.

\begin{figure}[!htb]
    \centering
    \subfloat[min-max normalization with shock]{
    \includegraphics[width=0.33\textwidth]{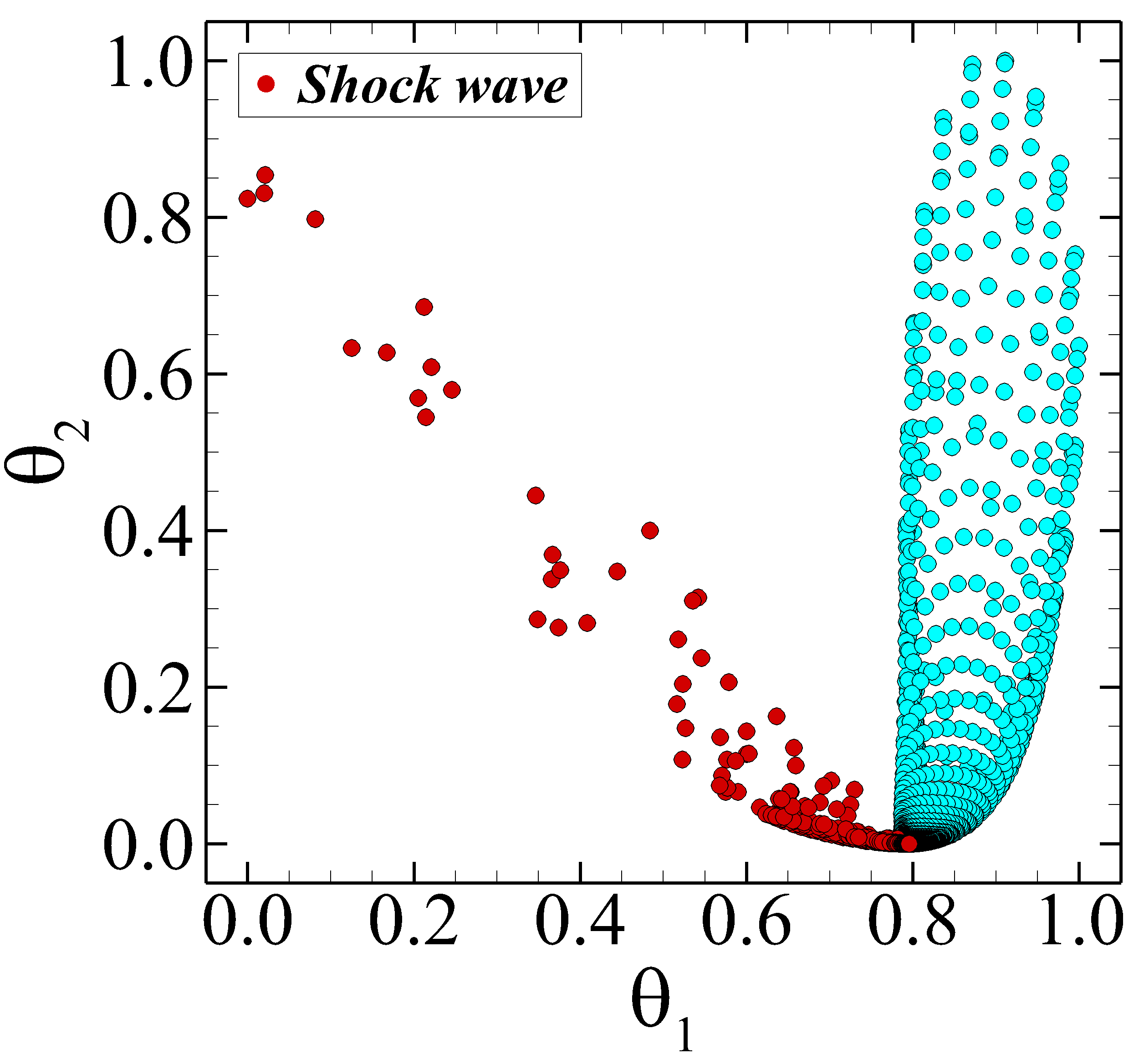} }
    \subfloat[min-max normalization without shock]{
    \includegraphics[width=0.33\textwidth]{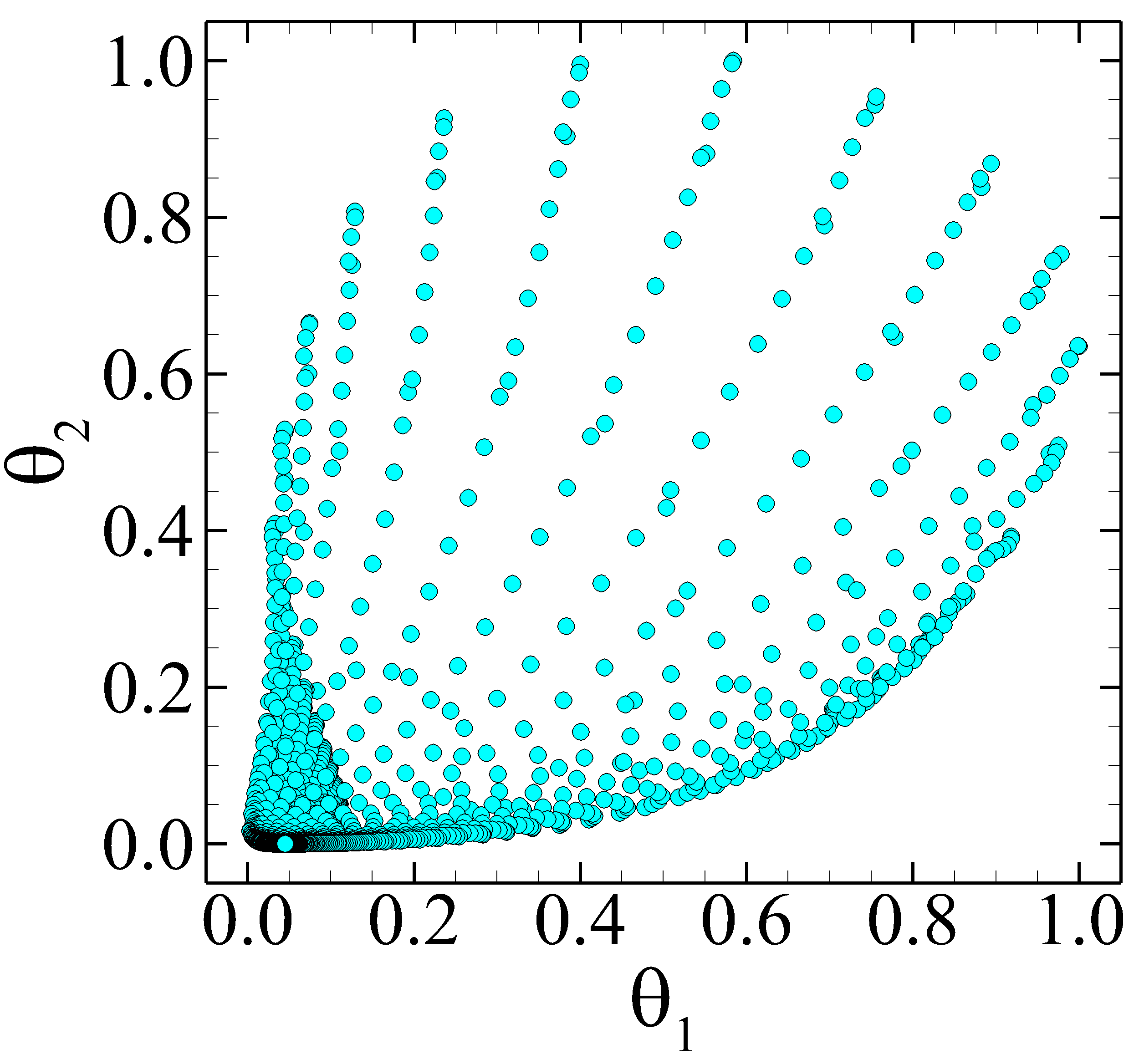} }
    \subfloat[local normalization]{
    \includegraphics[width=0.33\textwidth]{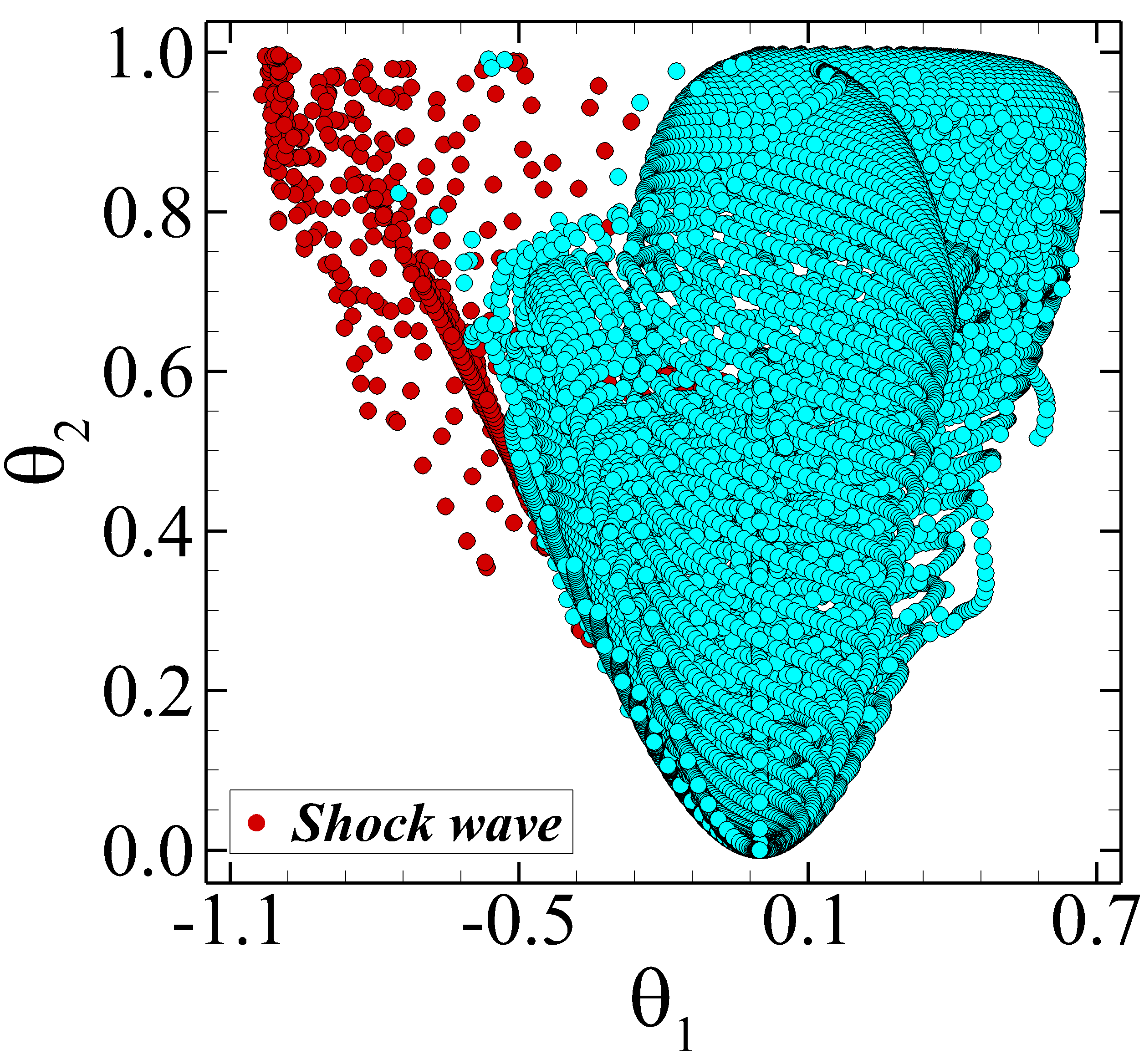} }
    \caption{Effects of shock waves on the distribution of input features. The red dots (\textcolor{red}{$\bullet$}) indicate the features in the shock wave region.}    
    \label{fig:sample_shock}
\end{figure}

The min-max normalization and local normalization are analyzed based on the distribution of the scalar invariants~$\theta_1$ and $\theta_2$ from the flow over the RAE2822 airfoils.
The scatter plots of the scalar invariants are presented in Fig.~\ref{fig:sample_shock}.
It can be seen that with the min-max normalization the shock wave would lead to most features narrowed down to the range of $[0.8, 1.0]$.
Such noticeable feature separations between the shock wave and the rest areas can cause sharp discontinuity in the scalar invariants.
In contrast, after removing the shock waves, the two scalar invariants~$\theta_1$ and $\theta_2$ have a relatively uniform distribution in the feature space as shown in Fig.~\ref{fig:sample_shock}(b), which has benefits to distinguish the subtle difference of input features.
It is noted that most invariants are distributed around $0$.
That is because parallel flows with small velocity gradients are dominant in the computational domain, and the invariants with large magnitudes only exist near the surface of airfoils.
As for the local normalization, the strategy can scale the features within the range of $[-1, 1]$ and simultaneously circumvent the feature collapse caused by the shock waves, as shown in Fig.~\ref{fig:sample_shock}(c).

We test the effects of the min-max normalization and the local normalization on the training performance for flows over the RAE2822 airfoil.
Generally, the min-max normalization leads to training divergence, likely due to the discontinuity in input features, while the local normalization can achieve good training performance in both accuracy and efficiency.
After removing the features in the shock wave regime, the min-max normalization can provide similar reductions in data misfit as the local normalization.
However, this strategy is very sensitive to the used criteria for shock wave detection since it requires identifying the shock wave area accurately.
In contrast, the local normalization has no such limitations and hence is adopted in this paper.

\subsection{Ensemble Kalman method for learning neural-network-based turbulence models}

The ensemble Kalman method is a statistical inference method based on Monte Carlo sampling, which has been widely used in various applications~\cite{zhang2020evaluation,schneider2022ensemble,zhang2022acoustic}.
This method randomly samples the weights of neural networks and uses the  statistics of the weights and model predictions to estimate the  gradient and Hessian of the cost function.
The cost function can be written as
$$ J= \| w^{n+1} - w^n  \|_\mathsf{P}+ \| \mathcal{H}[w^{n+1}] - \mathsf{y} \|_\mathsf{R},$$
where $w$ is the weight of the neural network, $n$ is the iteration index, $\mathsf{P}$ is the model error covariance, $\mathsf{R}$ is the observation error covariance, $\mathsf{y}$ is the observation data that obey the normal distribution with zero mean and variance of $\mathsf{R}$, and $\mathcal{H}$ is the model operator that maps the neural network weights to the observed quantities.
This method uses the ensemble of the realizations $W$ to estimate the sample mean~$\bar{W}$ and covariance~$\mathsf{P}$ as
\begin{equation}
\begin{aligned}
	\bar{W} &= \frac{1}{M} \sum_{m=1}^M w_{m} \text{,} \\
	\mathsf{P} &= \frac{1}{M-1} (W - \bar{W})(W-\bar{W})^\top \text{,}
\end{aligned}
\end{equation}
where $M$ is the sample size and $m$ is the sample index.
Based on the Gauss-Newton method, the weights require the first and second-order derivatives of the cost function to update the weights.
The ensemble Kalman method uses the statistics of these samples to estimate the derivative information~\cite {luo2015iterative}.
At the $n$~th iteration, each sample~$w_m$ is updated based on
\begin{equation}
    w_m^{n+1} = w_m^n + \mathsf{PH}^\top (\mathsf{HPH}^\top + \mathsf{R})^{-1} (\mathsf{y}_m^n - \mathsf{H}w_m^n) \text{,}
    \label{eq:enkf}
\end{equation}
where $\mathsf{H}$ is the tangent linear model operator.
The readers are referred to Ref.~\cite{zhang2022ensemble} for details of the ensemble-based turbulence modeling framework.

\begin{figure}[!htb]
    \centering
    \includegraphics[width=0.9\textwidth]{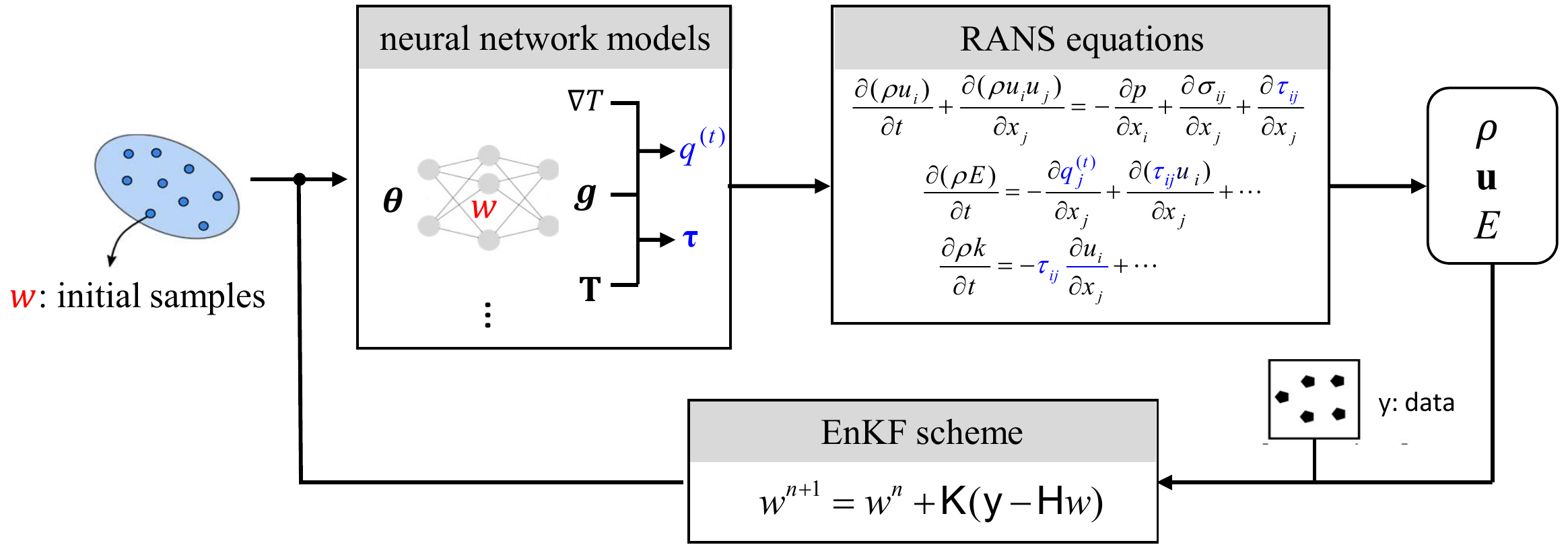}
    \caption{Schematic of ensemble-based training of tensor-basis neural network for transonic flows. 
    }
    \label{fig:schematic}
\end{figure}

The schematic of the machine learning framework for compressible flows is shown in Fig.~\ref{fig:schematic}.
Given the pre-trained weights~$w^0$ and the initial variance, we can draw an ensemble of samples for the neural network weights.
The neural network output, i.e., the tensor coefficient~$\mathbf{g}$, is combined with the tensor bases~$\mathbf{T}$ to construct the Reynolds stress tensors~$\boldsymbol{\tau}$.
Also, the coefficient~$\mathbf{g}$ is combined with the temperature gradient to form the turbulent heat flux~$q^{(t)}$.
The constructed Reynolds stress~$\boldsymbol{\tau}$ is used for the momentum equation, turbulent kinetic energy transport equation, and the energy equation, while the turbulent heat flux~$q^{(t)}$ is used for the energy equation.
Through coupling the neural network and the Reynolds stress, the flow prediction is obtained by solving the RANS equations.
Further the model prediction and training data are analyzed to update the weight of the neural network based on the ensemble Kalman method.

The detailed training procedure is formulated as follows.
\begin{itemize}
    \item Pre-training: Conventional weight initialization may lead to negative eddy viscosity, which may further cause the divergence of the solver.
    Hence, the neural network is pretrained to be a linear eddy viscosity model with $g^{1}=-0.09, g^{(2-10)}=0$.
    The samples of neural network weights are drawn around the pretrained weights, which ensure the obtained eddy viscosity close to the baseline.
    Moreover, the samples that lead to negative eddy viscosity are rejected and redrawn from the prescribed normal distribution.
    \item Feature extraction: The predicted flow field is used to construct the input features, i.e., the scalar invariants~$\mathbf{\theta}$ and the tensor bases~$\mathbf{T}$.
    Further, the invariants are normalized based on the local time scales as in Eq.~\eqref{eq:nl}.
    \item Coupling neural networks and the RANS equations: The input features are propagated to the $\mathbf{g}$ function through the neural network.
    The $\mathbf{g}$ function is further combined with tensor bases and temperature gradient to construct the Reynolds stress and turbulent heat flux, respectively.
    The constructed Reynolds stress and heat flux are used to close the RANS equations for estimating the mean flow field.
    \item Kalman correction:
    The flow prediction and the observation data are analyzed to update the weights of each neural network based on the ensemble Kalman method.
    Return to the second step until the maximum iteration is reached.
\end{itemize}

An in-house computational fluid dynamics (CFD) solver \cite{liu2019numerical,liu2018dynamic,Liu2020RSM-IDDES,wang2021iddes} is used to solve the RANS equations~\eqref{eq:NSEquations}. 
The governing equations are discretized with the cell-centered finite volume method on unstructured hybrid meshes composed of hexahedrons, prisms, tetrahedrons, and pyramids. 
The convective flux terms are computed with the second-order Roe discretization schemes~\cite{roe1986characteristic}, and the viscous flux terms are obtained by the reconstructed central scheme.
The LU-SGS relaxation-based implicit backward-Euler scheme is implemented for steady flow simulations. 
An adaptive local time-stepping method is developed to eliminate the adverse influence of some poor-quality grids on solution stability and convergence.
All calculations are performed with double precision on a High-performance computing (HPC) platform with 2.1GHz Intel(R) CPU, and the CFD code is parallelized by using a domain decomposition strategy with the
message passing interface (MPI) protocol. 
Nonblocking communications are used to overlap the computation with the communication and exploit possible performance gains.
The ensemble Kalman method is implemented in the publicly available DAFI code~\cite{strofer2021dafi}.

\section{Numerical results}
\label{sec:3-NumericalResults}

We present two cases to demonstrate the capability of the proposed framework in learning neural network-based turbulence models for transonic flows.
The test cases include 2D flows over the RAE2822 airfoil and 3D flows around the ONERA M6 wings, as summarized in Table~\ref{tab:DetailsSet}. 
The adiabatic wall condition is applied in this work, which is widely used as a thermal boundary condition~\cite{bendiksen2011review,roy2006review}. 
Specifically, we implement the adiabatic wall conditions in the energy equation Eq.~\eqref{eq:NSEquations}c by setting the gradient of temperature as zero at walls. 
The learned turbulence model and heat flux closure apply the solid wall boundary conditions for $k$ and $\omega$ by inheriting from the baseline $k$--$\omega$ SST model~\cite{blazek2015computational}.
We note that the learned models can also be applied for isothermal-wall cases by prescribing the wall temperature.
The wall boundary conditions for the turbulence model and the heat flux closure are consistent with the adiabatic-wall scenarios, which are independent of the wall temperature.
Such implementation strategies are commonly used to numerically investigate the hot or cold wall effects in aerospace applications~\cite{raje2021anisotropic,fluids4010037}.
The training data for the two cases are from the experimental measurements in  velocity and wall pressure, respectively.
The learned model for each case is tested in different flow conditions to demonstrate the predictive ability of the neural network-based model.

The ensemble Kalman method has good training efficiency since it is a second-order optimization with approximated gradient and Hessian information.
Specifically, the method draws an ensemble of samples, and the ensemble statistics are used to estimate the gradient and Hessian~\cite{luo2015iterative}.
For each sample, the neural network-based turbulence model will be coupled with the RANS equations to predict the flow fields.
The ensemble of cases can be run in a parallel manner without requiring communication.
Hence, the wall time is not significantly increased when these CFD cases are allocated to different CPUs.
Specifically, the computational cost for the RAE2822 case and the ONERA M6 case is about $232.3$ and $973.6$ core hours, respectively, which are mostly taken by the CFD calculations.
In the following, we present the detailed case setup and the corresponding training results for each case.

\begin{table}[!htb]
    \caption{\label{tab:DetailsSet} Details of the numerical settings 
    }
    \centering
    \begin{tabular}{ccccc}
    \hline
    \hline
    Cases  &  Geometry  &  Freestream conditions  & data~$\mathsf{y}$  & $\text{dim}(\mathsf{y})$\\
    \hline
    2D training case &  RAE 2822 airfoil  &  $Ma = 0.75, \alpha = 2.80^\circ, Re_C = 6.2 \times {10^6}$  &  $\mathbf{u}$ & $31$ \\
    2D test case (1) &  RAE 2822 airfoil  &  $Ma = 0.73, \alpha = 2.80^\circ, Re_C = 6.5 \times {10^6}$  &  $\mathbf{u}$ & $25$ \\
    2D test case (2) &  RAE 2822 airfoil  &  $Ma = 0.73, \alpha = 2.80^\circ, Re_C = 2.7 \times {10^6}$  & $\mathbf{u}$ & $28$\\
    3D training case &  ONERA M6 wing  &  $Ma = 0.84, \alpha = 5.06^\circ, Re_{MAC} = 1.17 \times {10^7}$  & $C_p$ & $52$ \\
    3D test case (1)  &  ONERA M6 wing  &  $Ma = 0.84, \alpha = 6.06^\circ, Re_{MAC} = 1.17 \times {10^7}$  & $C_p$ &  $52$ \\
    3D test case (2)  &  ONERA M6 wing  &  $Ma = 0.84, \alpha = 3.06^\circ, Re_{MAC} = 1.17 \times {10^7}$  & $C_p$ &  $52$ \\
    \hline
    \hline
    \end{tabular}
\end{table}

\subsection{Transonic flows over the RAE2822 airfoils}
    \label{sec:3.2-RAE2822} 
\begin{figure}[!htb]
    \centering
    \includegraphics[width=0.8\textwidth]{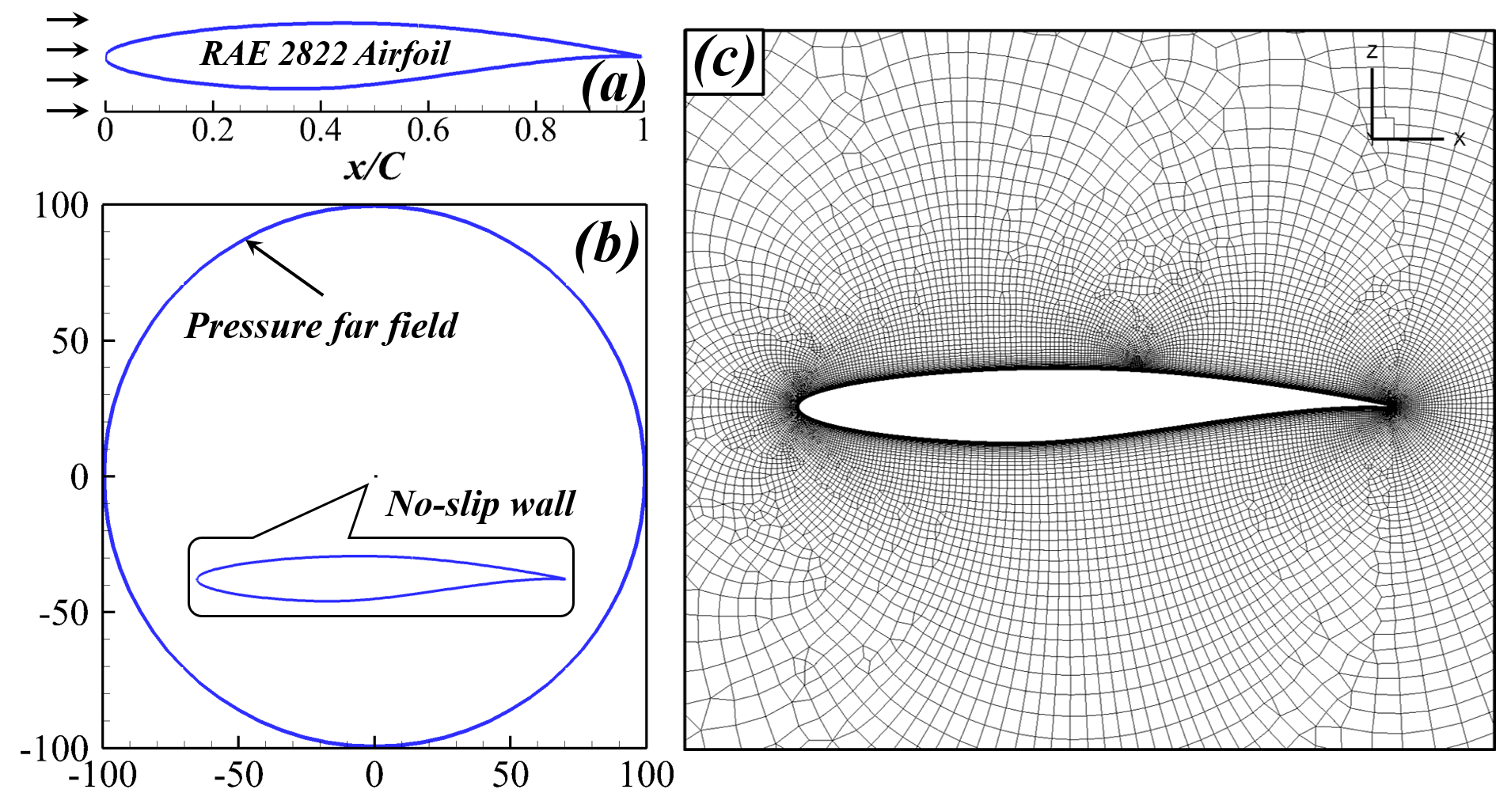}
    \caption{The computational setup for the RAE2822 case. (a) RAE 2822 airfoil; (b) Boundary conditions; (c) Zoomed view of the mesh near the airfoil. }
    \label{fig:RAE2822Grid}
\end{figure}

    The hybrid unstructured mesh is used for transonic flows over the RAE2822 airfoil as shown in Fig.~\ref{fig:RAE2822Grid}. 
    The mesh is chosen based on grid sensitivity tests, which contains $68,560$ vertices and $45,394$ cells. 
    Figure~\ref{fig:RAE2822Grid} (b) shows the schematic view of the computational domain. 
    The pressure far field (PFF) is a circle with radius of 100 chord lengths and center at the leading edge of the airfoil. 
    The adiabatic no-slip wall condition is employed on the airfoil surface. 
    Body-fitted O-type structured mesh is generated in the nearby region of the airfoil surface, and the triangle and quadrangle meshes are used in the rest of the computational domain. 
    The first cell height in the normal direction corresponds to $y^+ \approx 0.8$ with a growth rate of $1.15$ in the boundary layer.
    The training case is performed at the freestream Mach number of $0.75$, the angle of attack of $\alpha = 2.8^\circ$, and the Reynolds number of $Re_C = 6.2 \times 10^6$ based on the chord length~$C$.
    These flow conditions are set according to the case $10$ of the experimental measurements~\cite{cook1979aerofoil}.

    We use a neural network with two hidden layers and $5$ neurons per layer based on our sensitivity analysis. 
    The details of the sensitivity study to the neural network architecture are presented in Appendix~\ref{sec:Psnesitivity}.
    For this 2D case, the first three scalar invariants~$\theta^{(1-3)}$ are used as the inputs, and the first three tensor coefficients~$g^{(1-3)}$ are used as the outputs~\cite{hellsten2005new,wallin2000explicit}.
    The four velocity profiles, at $x/C=0.319$, $x/C=0.404$, $x/C=0.498$ and $x/C=0.574$, are used as training data.
    The relative standard deviation of the neural network weights is set as $0.3$, and the sample size is set as $32$ for this case.
    The relative observation error is set as $0.1 \%$, and the experimental data is the velocity measurements from the case $10$ of the experiments~\cite{cook1979aerofoil}.

\begin{figure}[!htb]
    \centering
    \subfloat[$k$--$\omega$ SST]{
    \includegraphics[width=0.46\textwidth]{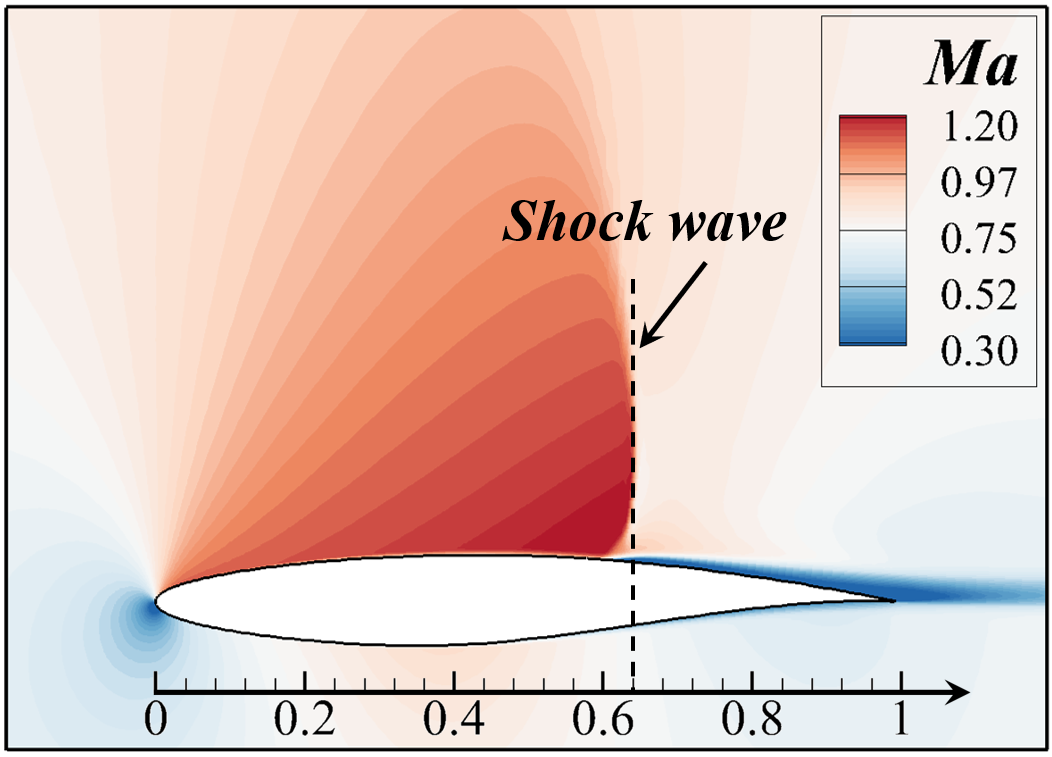} } 
    \hspace{5mm}
    \subfloat[Learned model]{
    \includegraphics[width=0.46\textwidth]{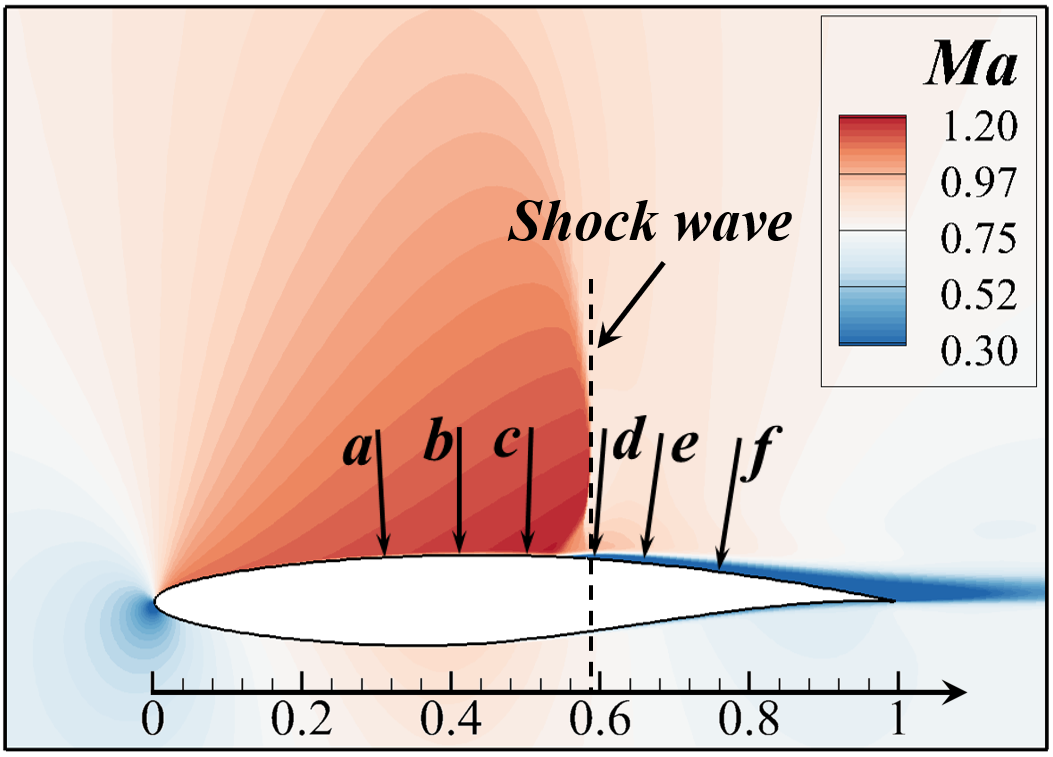}}
    \caption{Contour plots of Mach number with the baseline model (a)  and learned model (b) for the RAE2822 case.
    }
    \label{fig:RAE2822Flow}
\end{figure}

    The transonic flows over the RAE2822 airfoil involve shock waves, mean pressure gradient, and interaction between shock wave and boundary layer, which pose difficulties for turbulence modeling to capture such complex flows.
    Figure~\ref{fig:RAE2822Flow} presents the contour plots of Mach number with the learned model and the baseline model.
    We can see that the flows around the leading edge accelerate to the supersonic regime above the airfoil, and the supersonic flows are terminated by a shock wave leading to a supersonic pocket.
    At the root of the shock wave, the boundary layer thickens due to the strong adverse pressure gradient induced by the shock wave, which develops downstream and eventually forms a wake flow. % as shown in Figure~\ref{fig:RAE2822Flow}(b).
    The presence of the boundary layer increases the effective thickness of the airfoil, which would also affect the location of the shock wave according to the transonic similarity principles~\cite{bendiksen2011review}. 
    For this reason, it is challenging to predict well the flow in the boundary layer and the position of the shock wave.  
    It can be seen in Fig.~\ref{fig:RAE2822Flow} that the baseline model predicts the position of the shock waves at $x/C=0.645$.
    In contrast, the learned model predicts a relatively small supersonic region, and the shock wave occurs at a relatively upstream position, i.e., $x/C=0.586$.
    The experimental data of shock wave locations are not available for comparison in this case.

\begin{figure}[!htb]
    \centering
    \includegraphics[width=1.0\textwidth]{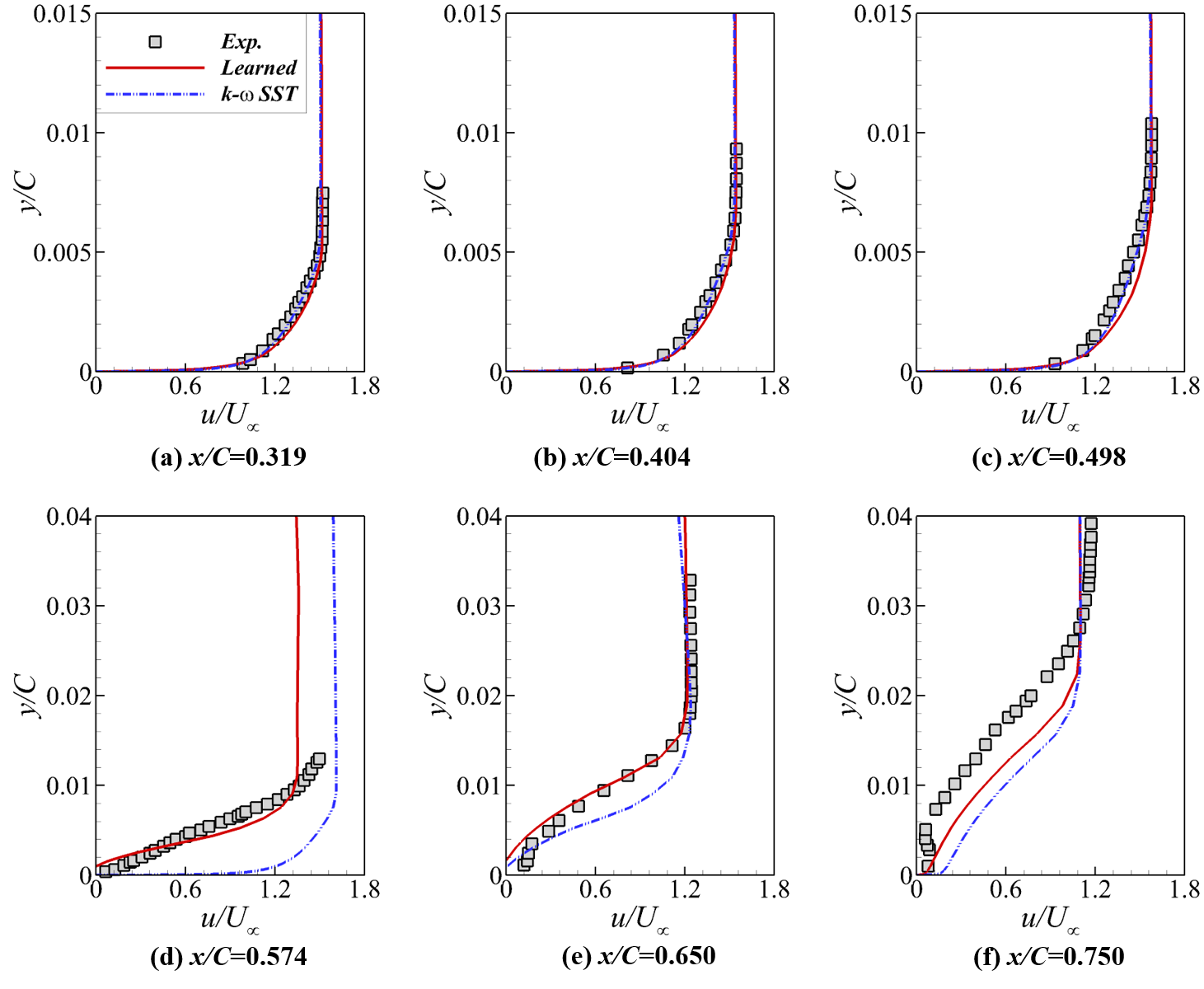}
    \caption{Plots of the velocity profiles with the comparison among the baseline model, the learned model, and the experimental data for the RAE2822 case. 
    }
    \label{fig:RAE2822VelProfile}
\end{figure}

    The learned model can significantly improve the estimation of velocity compared to the baseline model. 
    It can be seen from Fig.~\ref{fig:RAE2822VelProfile}, which presents the velocity profiles at different locations on the upper surface of the RAE2822 airfoil. 
    The locations are distributed on both sides of the shock wave as indicated in Figure~\ref{fig:RAE2822Flow} (b).
    The three locations of $a, b$, and $c$ are at the upstream of the shock wave, and the locations of $d, e$, and $f$ are at the downstream of the shock wave.
    At the locations of $a, b$, and $c$, the characteristics of the velocity profiles are similar, and the boundary layer thickens slowly.
    In contrast, at the locations of $d, e$, and $f$, the boundary layer thickens rapidly due to the adverse pressure gradient.
    The maximum value of the adverse pressure gradient is located at the root of the shock wave, which is between the locations of $x/C=0.498$ and $x/C=0.574$.
    Such strong adverse pressure gradients result in the thickness of the boundary layer $\delta$ rapidly increasing from $\delta =0.0055C$ at the location $x/C=0.498$ to $\delta =0.0083C$ at the location $x/C=0.574$. 
    When the boundary layer develops downstream to the location $x/C=0.750$, the thickness can reach $\delta =0.0335C$. 
    From the comparison, the learned model can better predict the velocities than the baseline model, particularly at the locations of $x/C=0.574$ and $x/C=0.650$.

\begin{figure}[!htb]
    \centering
    \subfloat[pressure coefficients]{
    \includegraphics[width=0.33\textwidth]{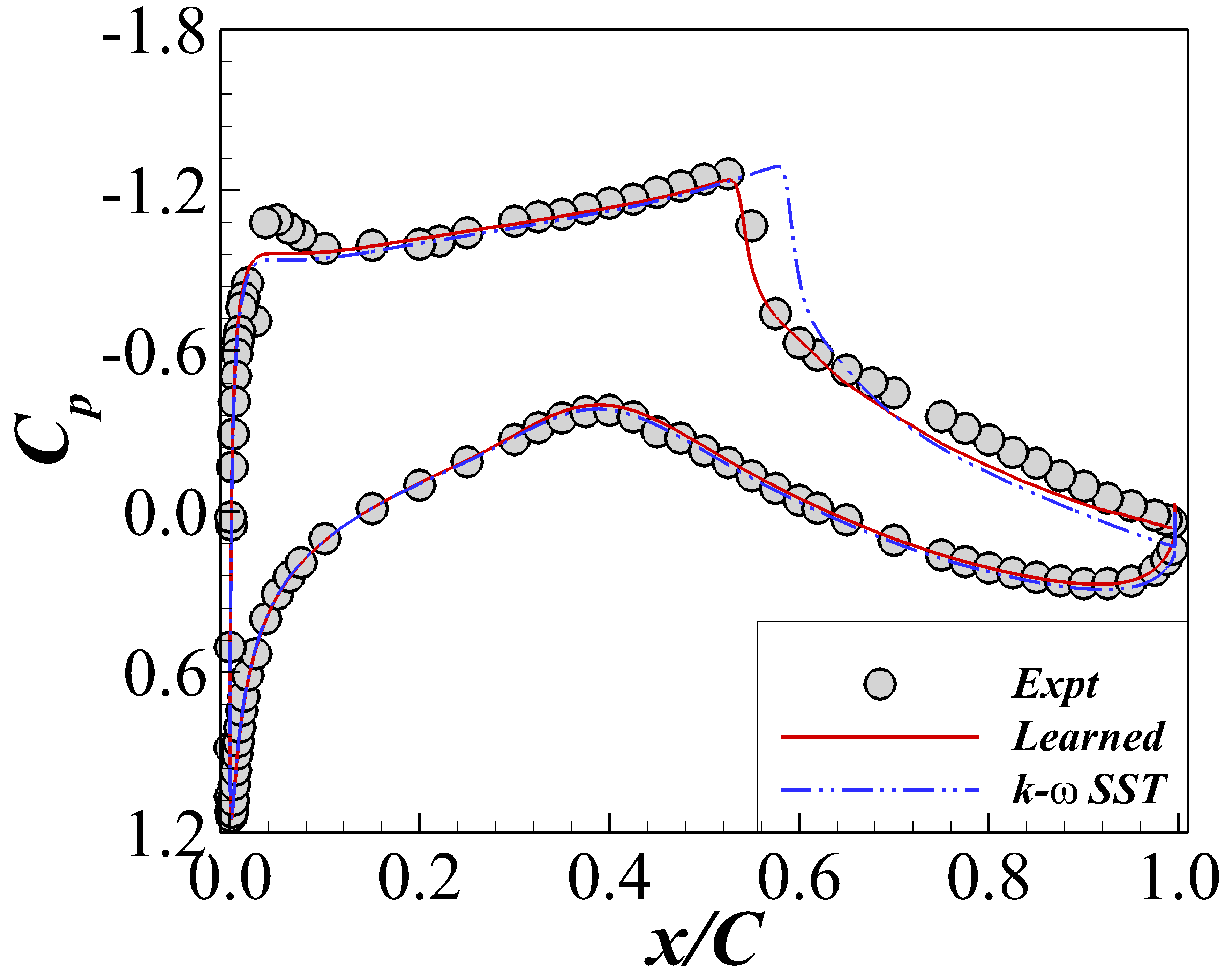} }
    \subfloat[friction coefficients]{
    \includegraphics[width=0.33\textwidth]{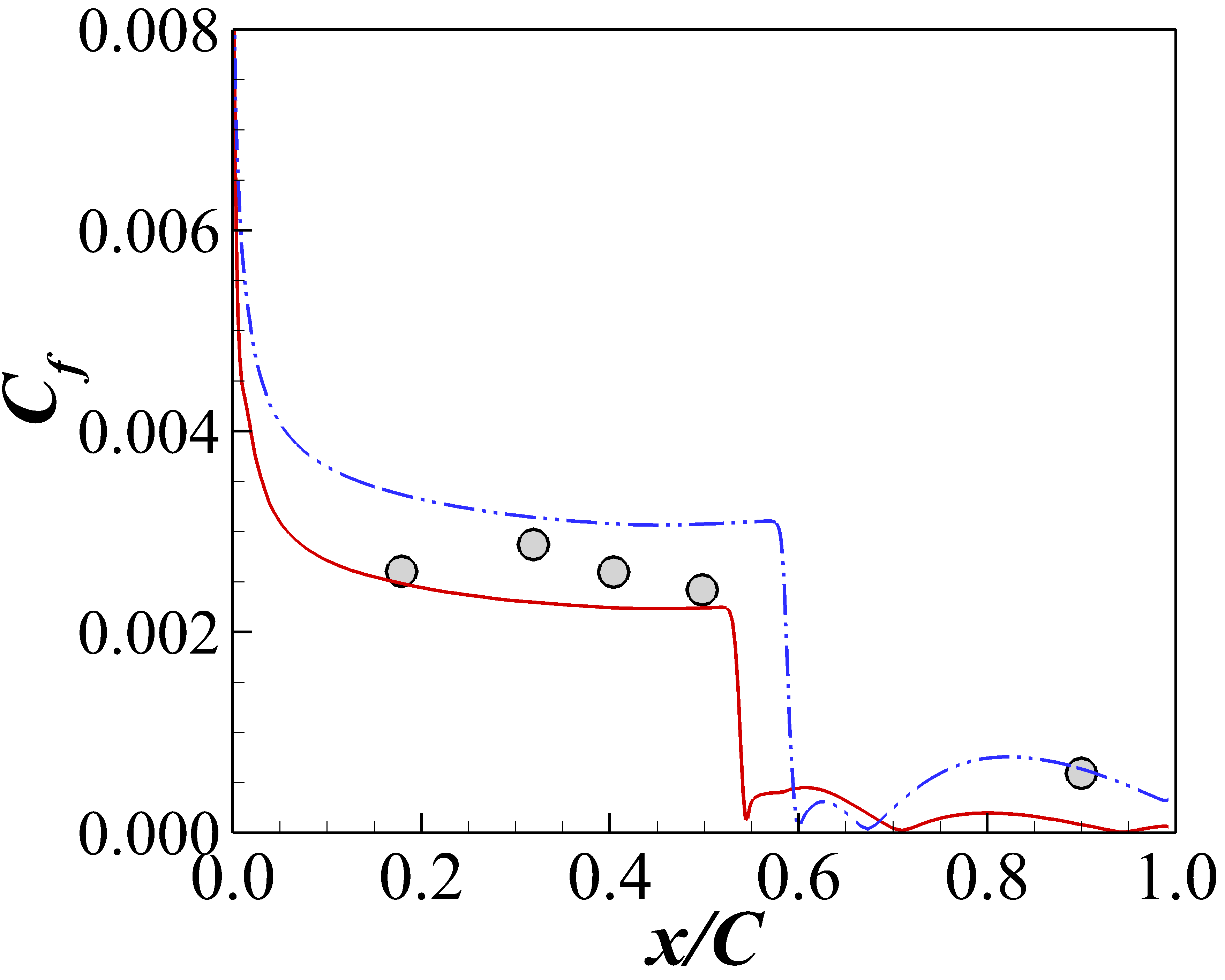} }
    \subfloat[wall temperature]{
    \includegraphics[width=0.33\textwidth]{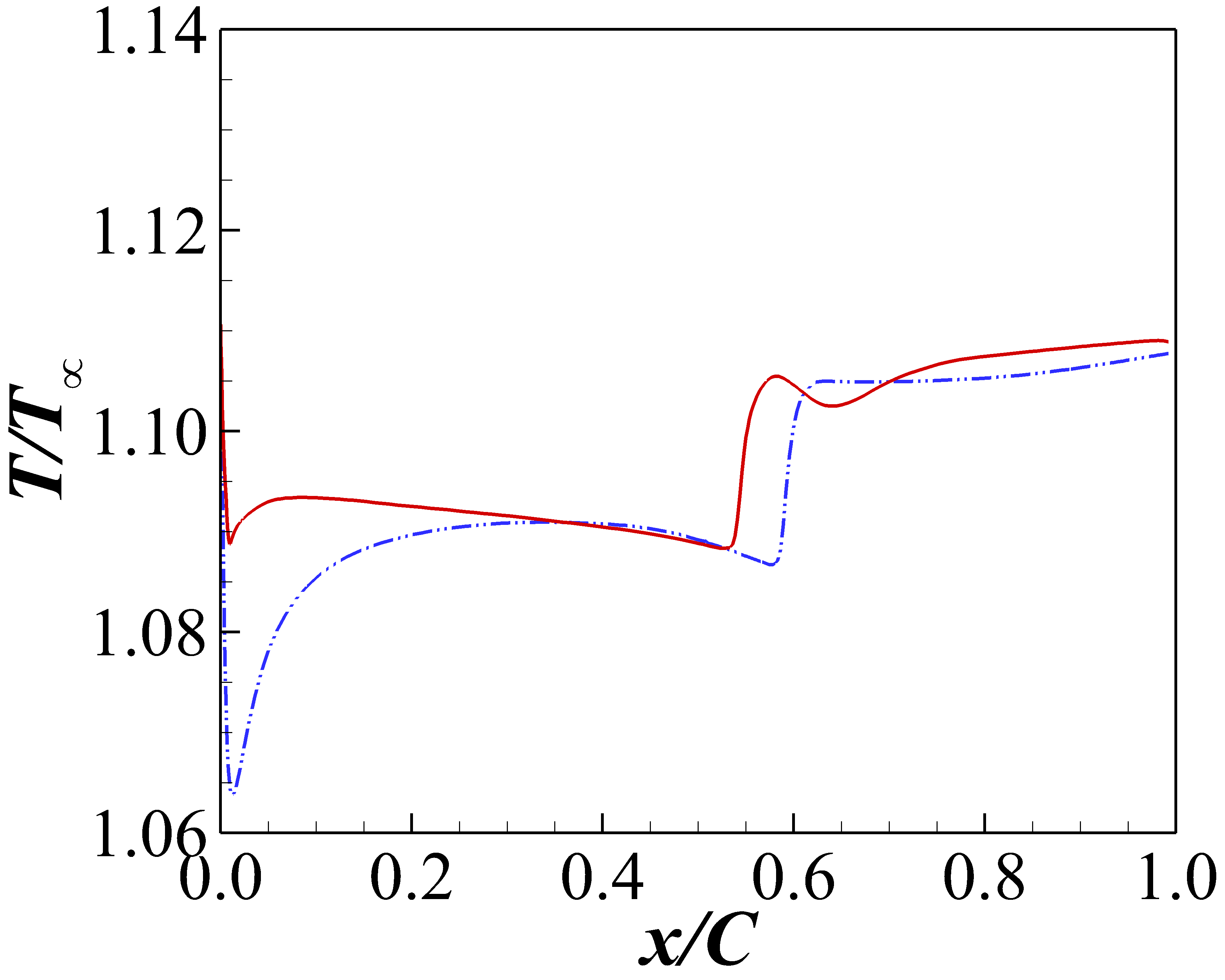} }
    \caption{Plots of the predicted pressure coefficients, friction coefficients, and wall temperature with comparison to the experimental data, for the RAE2822 case.}
    \label{fig:RAE2822CpandCf}
\end{figure}

\begin{figure}[!htb]
    \centering
    \subfloat[$g^{(1)}$-distribution]{
    \includegraphics[width=0.46\textwidth]{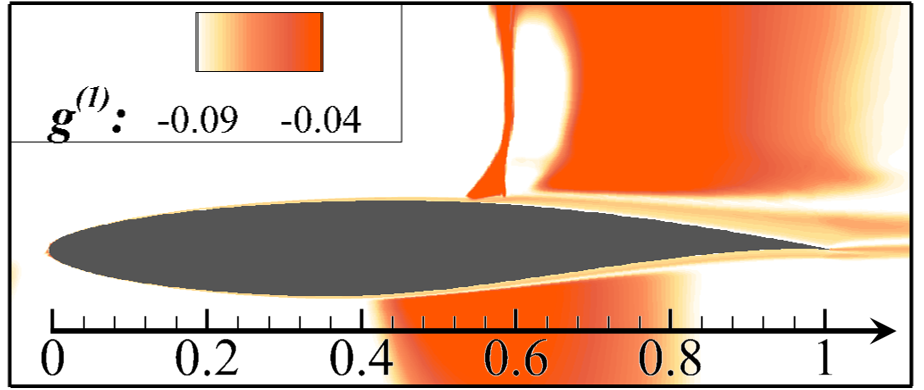} }
    \hspace{3mm}
    \subfloat[$g^{(2)}$-distribution]{
    \includegraphics[width=0.46\textwidth]{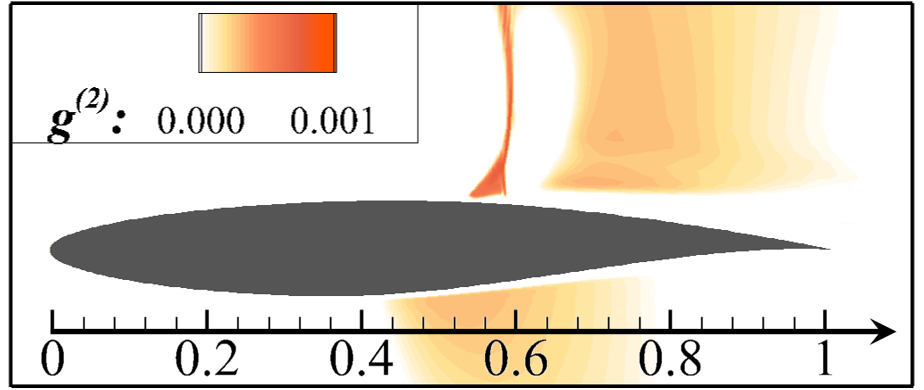} }
    \caption{Plots of $g^{(1)}$ and $g^{(2)}$ coefficients for the RAE2822 case. }
    \label{fig:RAE2822Gfunction}
\end{figure}

    The learned model is able to improve the prediction of the wall pressure, friction coefficient, and wall temperature from the baseline model.
    It can be seen clearly in Fig.~\ref{fig:RAE2822CpandCf}, which provides the pressure and friction coefficient in comparison to the experimental measurements.
    The learned model provides better agreements with the experimental data in the pressure coefficient, particularly near the shock wave areas, compared to the baseline $k$--$\omega$ SST model.
    This is because the learned model decreases the Reynolds stress in the boundary layer downstream of the shock wave, which leads to the boundary layer thickening.
    The thickened boundary layer reduces the effective camber of the airfoil, which drives the shock wave upstream and changes the pressure distribution in the rear of the airfoil.
    Evidence can be found in Figure~\ref{fig:RAE2822Gfunction}(a), which shows the distribution of the $g^{(1)}$ and $g^{(2)}$ coefficients near the airfoil. 
    The $g^{(1)}$ coefficient in the boundary layer downstream is around $-0.065$, which has a smaller magnitude than the baseline value $-0.09$. 
    That means that the learned model reduces the Reynolds stress compared to $k$--$\omega$ SST model based on Eq.~\eqref{eq:ReStress}. 
    Figure~\ref{fig:RAE2822Gfunction} (b) presents the $g^{(2)}$ distribution in the region near the airfoil.
    The amplitude of $g^{(2)}$ coefficient is close to zero in most of the areas with a maximum value of $1.2\times10^{-3}$, which indicates that the Reynolds stresses are dominated by the linear part of the learned model for the RAE2822 case.
    Figure~\ref{fig:RAE2822CpandCf} (b) presents the distribution of the friction coefficients, which also shows that the learned model makes better predictions than the baseline model.
    Figure~\ref{fig:RAE2822CpandCf} (c) presents the distribution of the wall temperature on the upper surface. 
    Due to the strong compressibility effect, the flow temperature will increase after passing through the shock wave. 
    Consequently, a small jump in wall temperature can be observed at the root of the shock wave ($x/C \approx 0.59$) for both the learned and the baseline models.
    Moreover, the learned model predicts a small reduction in the wall temperature after the shock wave.

\begin{figure}[!htb]
    \centering
    \includegraphics[width=1.0\textwidth]{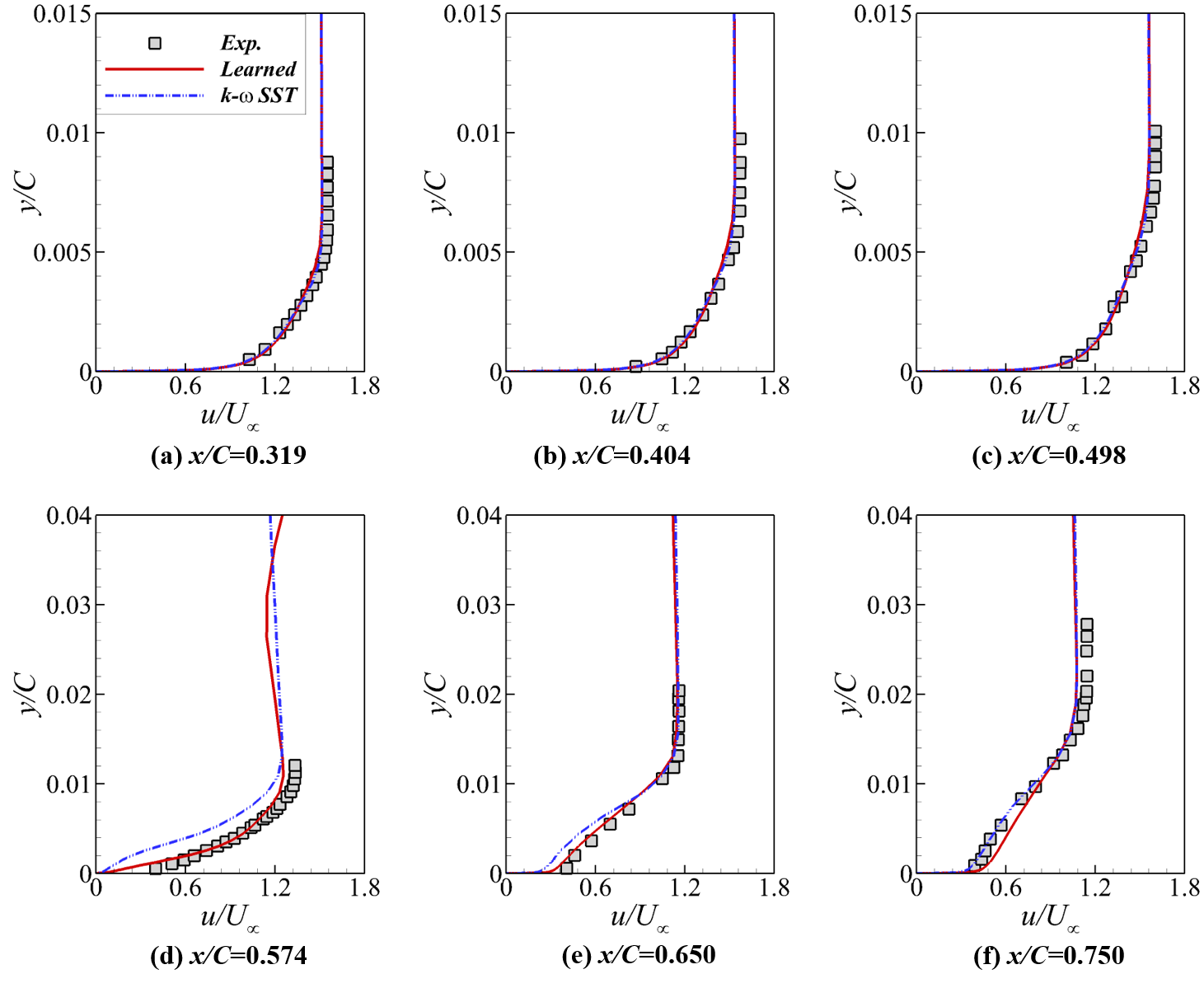}
    \caption{Velocity results of generalization tests with comparison among the baseline model, the learned model, and experimental data, for the RAE2822 case.
    }
    \label{fig:RAE2822VelProfileFcastCase09}
\end{figure}

    The generalization test shows that the trained neural network-based turbulence model is capable of predicting flow fields at different flow conditions.
    We test the learned turbulence model to predict the flow over the RAE2822 airfoils at two flow conditions.
    One has a similar Reynolds number $Re_C = 6.5\times 10^6$ to the training case, while the other has a much lower Reynolds number $Re_C = 2.7\times 10^6$.
    Both cases have the Mach number~$0.73$ and the angle of attack~$2.80^\circ$, which are similar to the training case.
    The two cases are also known as Case 9 and Case 12~\cite{cook1979aerofoil}, respectively.
    The transonic flows are very sensitive to the inflow conditions~\cite{witteveen2009uncertainty,pisaroni2017continuation} for the RAE2822 case. 
    Small variations can lead to noticeable changes in shock wave locations.
    Specifically, for the 2D test case (1), the flow with the learned model provides a relatively upstream location of the shock wave at $x/C=0.563$ compared to the $x/C=0.586$ of training case, and for the 2D test case (2) the predicted shock wave is located further upstream at $x/C=0.498$. 
    The angle of attack is given by~$\alpha = 3.19^\circ$ in the experiments, while it is recommended to be set as~$\alpha = 2.8^\circ$ in numerical simulations to consider the wind-tunnel-wall interference~\cite{cook1979aerofoil}. 
    Our results show that the learned model can achieve predictive improvements in the velocity and wall pressure for both the test cases.
    Only the velocity plots for the 2D test case (1) are shown in Fig.~\ref{fig:RAE2822VelProfileFcastCase09} for brevity.
    It can be seen that the velocity prediction with the learned model can provide noticeable improvements over the baseline $k$--$\omega$ SST model.
    Particularly, in the locations of (d) and (e) behind the shock waves, the learned model achieves significant improvements from the baseline model.
    At location (f), there are only slight differences between the two models, both of which have satisfactory agreements with the experimental data.

The error between the prediction and experimental measurements is quantified based on
\begin{equation}
    \text{error} = \frac{\| q - q^\text{exp}\|}{\| q^\text{exp} \|} \text{.}
\end{equation}
The error estimate for the training and test cases are summarized in Table~\ref{tab:ErrorEstmt2822}.
It can be seen that the learned model achieves better prediction accuracy compared to the baseline model.
In the training case, the error in terms of velocity is reduced significantly.
Specifically, the velocity error is reduced from $14.819 \%$ to $7.868 \%$,
while the error in $C_p$ and $C_f$ is reduced from $12.582\%$ to $7.701\%$  and from $21.742 \%$ to $16.531\%$, respectively. 
For the 2D test case (1) with $Ma=0.73, \alpha=2.80^\circ, Re_C=6.2 \times 10^6$, the velocity prediction is significantly improved from $9.343\%$ to $6.309\%$ with the learned model.
The pressure coefficient~$C_p$ is improved slightly from $8.1\%$ to $6.95 \%$, while the friction coefficient~$C_f$ is varied from $21.568 \%$ to $23.3 \%$.
Note that only velocity data is used for model training in this case.
Hence, the prediction improvement in velocity is expected, while the improvement in pressure and friction coefficients can not be guaranteed unless these data are also used for model training.
For the 2D test case (2) with $Ma=0.73, \alpha=2.80^\circ, Re_C=2.7 \times 10^6$, the velocity prediction error is reduced from $9.856\%$ to $6.769\%$, and the prediction error in wall pressure is reduced from $11.020\%$ to $7.649\%$. 
The friction coefficient measurements are not available in this test case.

\begin{table}[!htb]
    \caption{\label{tab:ErrorEstmt2822} Summary of the prediction error in the velocity, pressure coefficient, and friction coefficient for the RAE2822 case. 
    }
    \centering
    \begin{tabular}{ccccc}
    \hline
    \hline
    \multirow{2}{*}{Cases}          & \multirow{2}{*}{Turbulence models}  & \multicolumn{3}{c}{Error estimation} \\ \cline{3-5} 
                                    &       & Velocity  & $C_p$     & $C_f$       \\
    \hline
    \multirow{2}{*}{Training case}  & $k$--$\omega$ SST   & $14.819 \%$ & $12.582 \%$ & $21.742 \%$  \\  % Expt = 196.3696502
                                    & learned model       & $ 7.868 \%$ & $ 7.701 \%$ & $16.531 \%$  \\
    \multirow{2}{*}{Test case (1)}      & $k$--$\omega$ SST   & $ 9.343 \%$ & $ 8.100 \%$ & $21.568 \%$  \\  % Expt = 160.8838424169 
                                    & learned model       & $ 6.309 \%$ & $ 6.950 \%$ & $23.300 \%$  \\
    \multirow{2}{*}{Test case (2)}      & $k$--$\omega$ SST   & $ 9.856 \%$ & $ 11.020\%$ & ---$^a$  \\  % Expt = 160.8838424169 
                                    & learned model       & $ 6.769 \%$ & $ 7.649 \%$ & ---$^a$  \\
    \hline
    \hline
    \multicolumn{5}{l}{$^a$ The $C_f$ measurements of test case (2) are not available.} \\
    \end{tabular}
\end{table}

\subsection{Transonic flows around 3D ONERA M6 wings}
    \label{sec:3.3-M6Wing} 
\begin{figure}[!htb]
    \centering
    \includegraphics[width=0.7\textwidth]{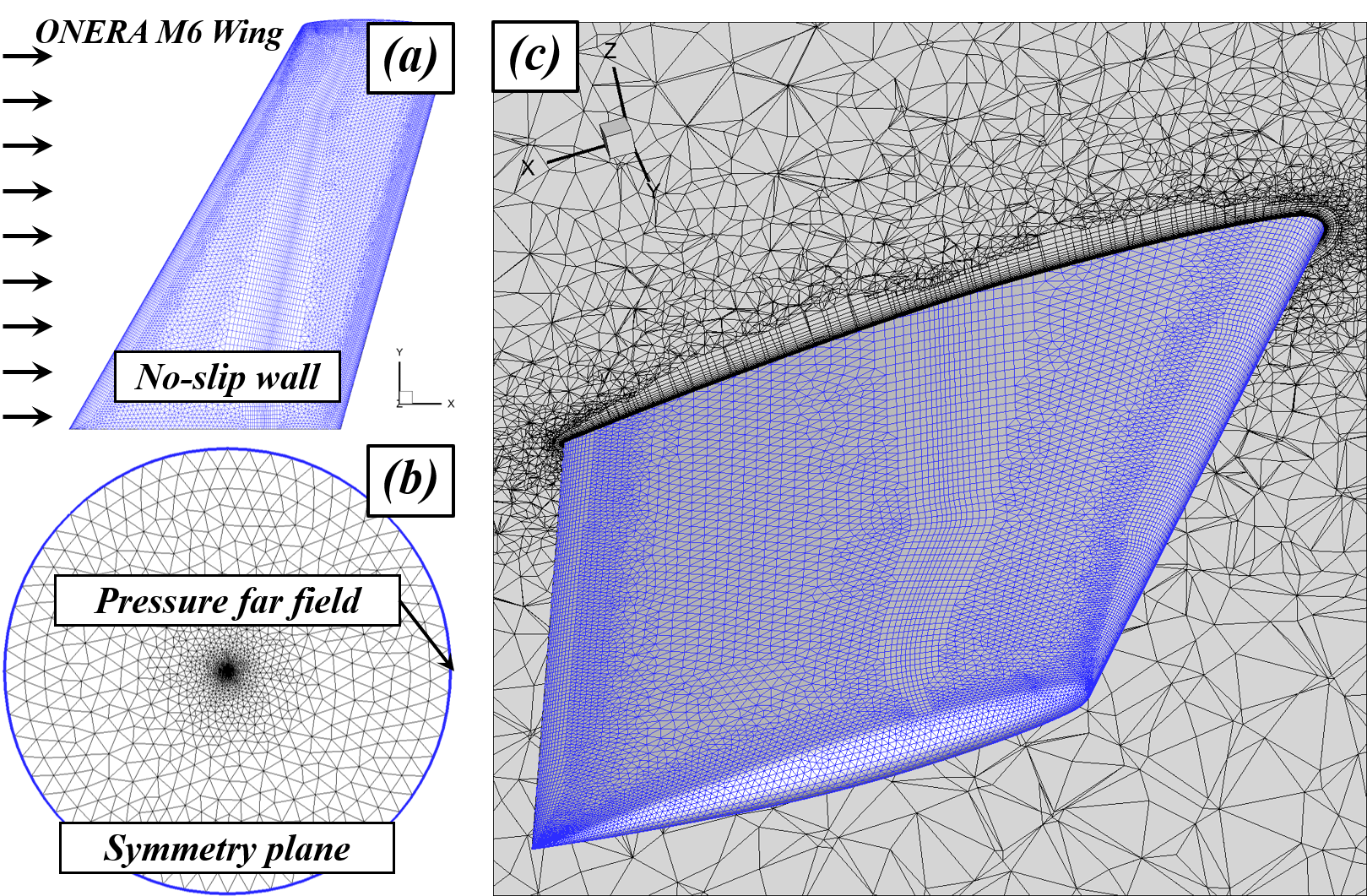}
    \caption{Computational setup for flows over ONERA M6 wings. (a) ONERA M6 wing; (b) boundary conditions; (c) mesh grids near the wing.}
    \label{fig:M6Mesh}
\end{figure}

    Flows over the ONERA M6 wings are typical external transonic flows, which have been widely used for numerical validation and turbulence model assessments. 
    The flows include shock-induced separation, especially at a relatively high angle of attack, i.e., $6.06^\circ$, which poses a challenge for turbulence modeling.
    Numerical simulations are performed on a hybrid unstructured mesh, as shown in Fig.~\ref{fig:M6Mesh}.
    The mesh employs a pressure far-field boundary condition at a distance of about 200 chord lengths, a symmetry boundary condition at the inboard wing side, and an adiabatic no-slip wall condition on the wing surface.
    The mesh contains $3,627,926$ cells and $1,602,721$ points.
    The grid points on the wing surface are clustered near the presumed shock wave positions and the wing tip to well capture the shock wave and shock-induced separations, respectively %, where shock-induced separations occur.
    The dimensionless height of the first cell satisfies $y^+ < 0.8$, and the growth rate of the mesh cells is around $1.2$ in the boundary layer.
    The training cases are conducted at Mach number $Ma=0.84$, angle of attack $\alpha=5.06^\circ$, and Reynolds number $Re_{MAC}=1.17 \times 10^7$ based on the mean aerodynamic chord (MAC)~\cite{schmitt1979pressure}. The mean aerodynamic chord is $0.64607$ m.

    In the ONERA M6 case, we employ a neural network with two hidden layers and $10$ neurons per layer based on our sensitivity analysis.
    The details of the sensitivity study to the neural network architecture are presented in Appendix~\ref{sec:Psnesitivity}.
    The six scalar invariants~$\theta^{(1-6)}$ are used as the inputs, and the five tensor coefficients~$g^{(1-5)}$ are used as the outputs based on the minimal representation principle~\cite{fu2011minimal}.
    The relative standard deviation of the neural network weights is set as $0.2$, and the sample size is set as $15$ for this case.
    The relative observation error is set as $0.1 \%$, and the observation data is the wall pressure distribution from the experiments~\cite{schmitt1979pressure}.
    The pressure measurements at four sections, i.e., $\eta =$ 0.44, 0.65, 0.90, and 0.99, are used as training data. 
    Herein, $\eta$ indicates the spanwise location, defined as $\eta=y/b$, where $y$ is the spanwise coordinate, and $b$ is the spanwise length.

\begin{figure}[!htb]
    \centering
    \includegraphics[width=0.95\textwidth]{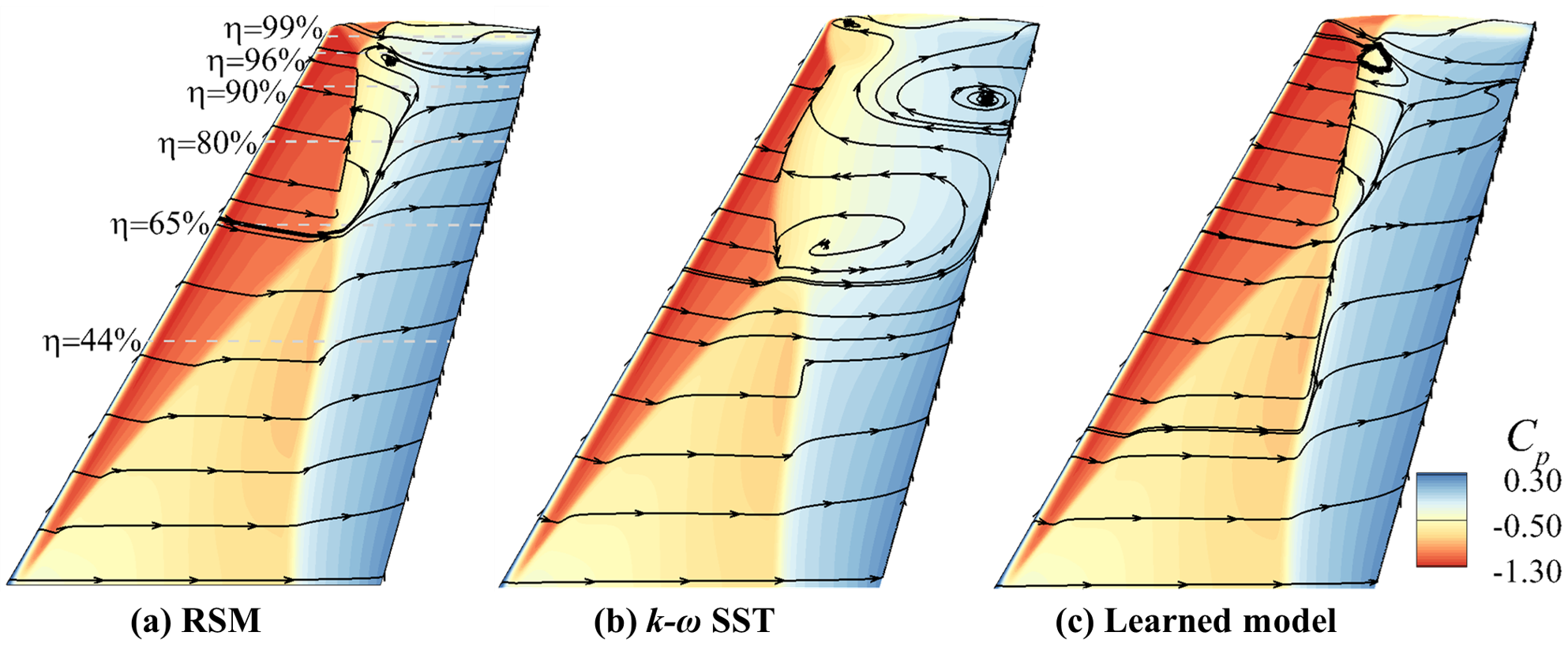}
    \caption{Plots of surface pressure and streamline with the Reynolds stress transport model, baseline model, and learned model, for the ONERA M6 case.}
    \label{fig:M6Flow}
\end{figure}

    The learned model can better estimate the flow separation than the baseline model.
    We use the prediction results of the Reynolds stress transport model in the previous study~\cite{liu2021numerical} as a reference. 
    Figure~\ref{fig:M6Flow} shows the surface pressure coefficient and the distribution of streamlines with comparison among the Reynolds stress transport model (RSM), $k-\omega$ SST model, and the learned model.
    From the pressure coefficient distributions, the different turbulence models predict similar development of the shock waves.
    The morphology of the shock waves is characterized by the formation of an $\lambda$-like shape on the upper wing. 
    That is, two shock waves are formed from the wing root and gradually approach the wing tip, merging into a stronger shock wave.
    As can be seen from Figure~\ref{fig:M6Flow}(b), the $k-\omega$ SST model predicts that the two shock waves on the upper surface of wings are merged at approximately $\eta = 0.55$, while the Reynolds stress transport model and the learned model predict the shocks merging at the spanwise $\eta = 0.65$. 
    Moreover, the predicted shock-induced flow separation zones are significantly different with the three turbulence models. 
    The $k-\omega$ SST model shows a large separation area after the shock wave, which develops from the root of the shock wave to the trailing edge of the wing. 
    In contrast, the learned model predicts similar results to the Reynolds stress transport model, with a relatively small shock-induced separation zone.
    Specifically, the Reynolds stress transport model, $k-\omega$ SST model, and the learned model predict the flow separation covering the range of $x_{RSM} / L_{MAC} \in [1.169, 1.323]$, $x_{k\omega} / L_{MAC} \in [1.091, 1.729]$, $x_{ML} / L_{MAC} \in [1.182, 1.349]$ in the streamwise direction, respectively.

\begin{figure}[!htb]
    \centering
    \includegraphics[width=1.0\textwidth]{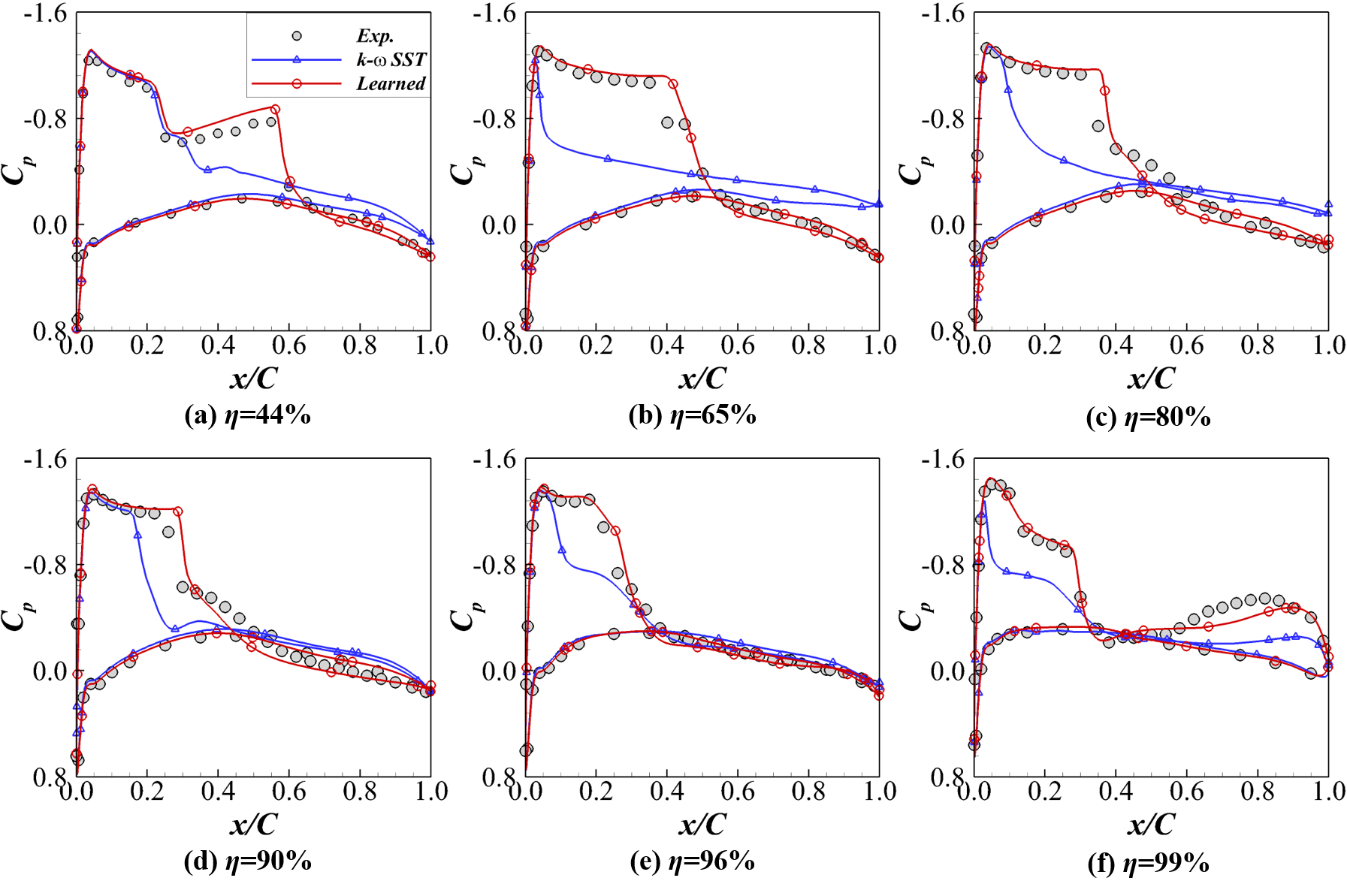}
    \caption{Plots of wall pressure coefficients with comparison among the baseline model, the learned model, and experimental data, for the ONERA M6 case.}
    \label{fig:M6Cp}
\end{figure}

    The learned model can provide improved predictions in terms of the pressure coefficient compared to the baseline model.
    It is supported by Fig.~\ref{fig:M6Cp}, which shows the pressure coefficient distribution at different spanwise positions on the wing surface in comparison to the experimental data.
    The position of the span-wise section is indicated in Figure~\ref{fig:M6Flow}(a), i.e., $\eta = 0.44, 0.65, 0.80, 0.90, 0.96, 0.99$ from the wing root to the wing tip.
    It can be seen that the pressure coefficients predicted by the baseline $k-\omega$ SST model have significant discrepancies from experimental data. 
    In contrast, the learned model is able to predict the pressure coefficient in good agreement with the experimental measurements in different spanwise sections.
    This is due to the fact that it accurately predicts the location of the $\lambda$-type shock waves, as shown in Fig.~\ref{fig:M6Flow}.

\begin{figure}[!htb]
    \centering
    \subfloat[$k$--$\omega$ SST]{
    \includegraphics[width=0.46\textwidth]{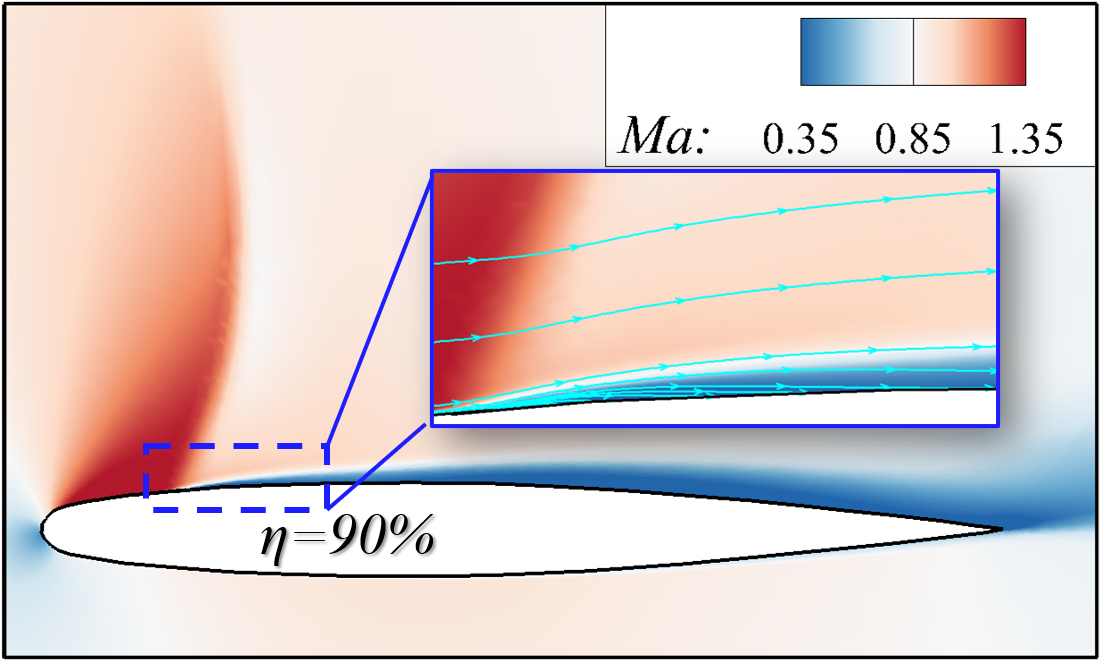} }
    \subfloat[ learned model]{
    \includegraphics[width=0.46\textwidth]{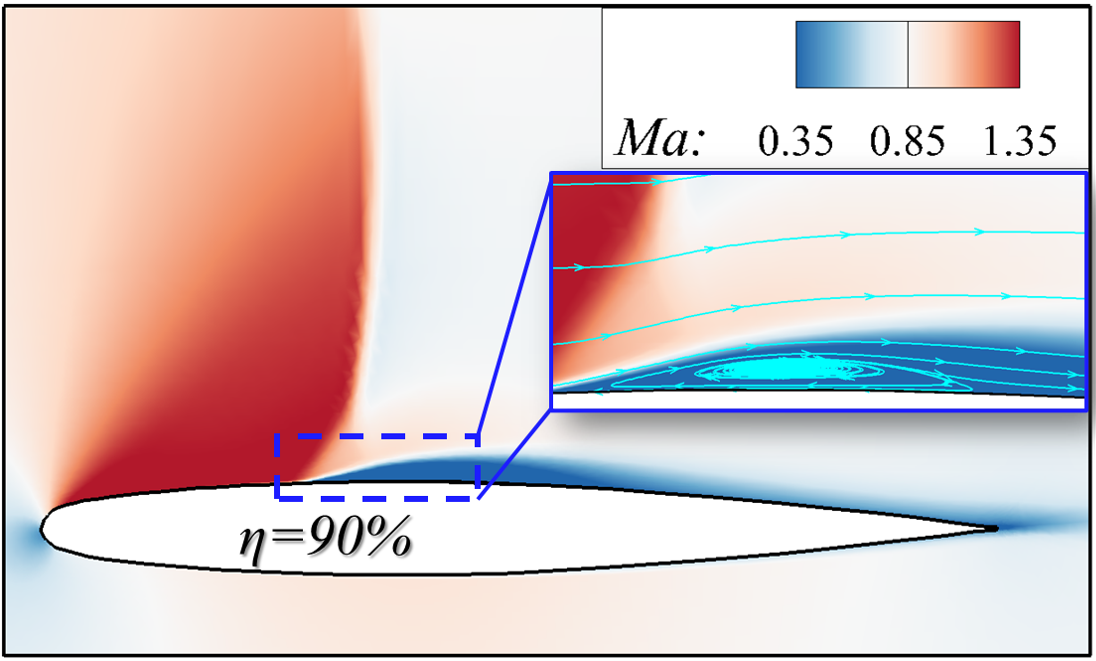} }
    \\
    \subfloat[$g^{(1)}$-distribution]{
    \includegraphics[width=0.46\textwidth]{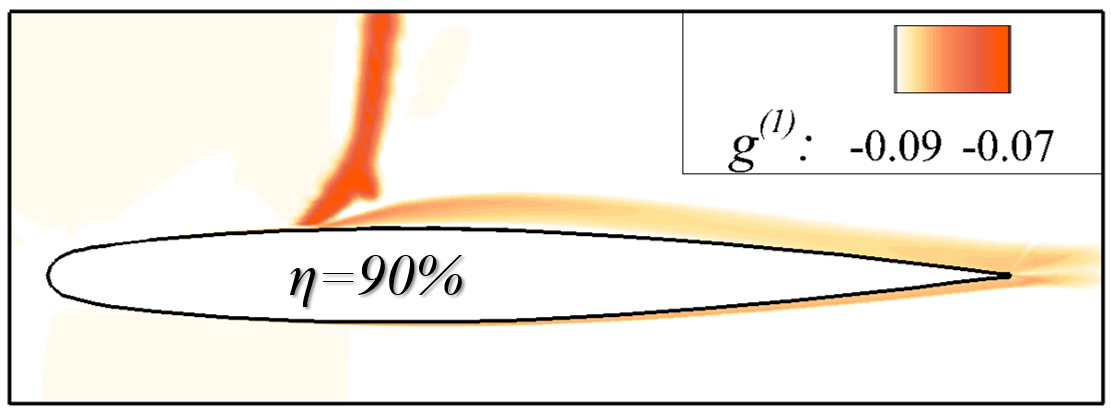} }
    \subfloat[$g^{(2)}$-distribution]{
    \includegraphics[width=0.46\textwidth]{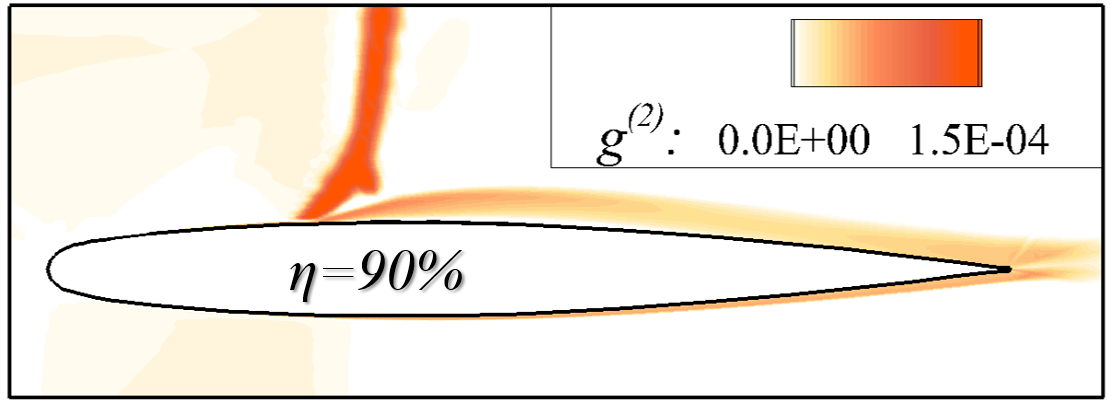} }
    \caption{Contour plots of the Mach number, $g^{(1)}$ and $g^{(2)}$ coefficients at the cutoff plane of $\eta = 90\%$ for the ONERA M6 case. 
    }
    \label{fig:M6Gfunction}
\end{figure}

    The predictive improvement in the surface pressure with the learned model is likely due to the reduction of Reynolds shear stress downstream of the shock wave.
    Figure~\ref{fig:M6Gfunction} shows the contour plots of Mach number, $g^{(1)}$ and $g^{(2)}$ coefficients obtained by the learned model on a cutoff plane $\eta = 0.90$, where the flow separation is the most extensive. 
    One can see that the supersonic pocket and the shock-induced separations on the suction side are resolved well. 
    In this separation region, the learned model reduces the magnitude of  $g^{(1)}$ coefficient and further the Reynolds shear stress based on Eq.~\eqref{eq:ReStress}. 
    The reduction of the shear stress weakens the momentum transport within the boundary layer, which leads to the thickening of the boundary layer downstream of the shock wave. 
    In addition, the enlarged separation zone results in a circular-arc bump that would hold the shock wave at a relative downstream position compared to the baseline model.
    Recall that $g^{(2)}$ is the coefficient of the second tensor basis, and the distribution of $g^{(2)}$ coefficient is shown in Figure~\ref{fig:M6Gfunction}(d). 
    The coefficients~$g^{(3-5)}$ have similar patterns as the $g^{(2)}$ with very small magnitudes, and hence their plots are omitted for brevity.
    The relatively large amplitude is mainly in the separation and wake regions, with a maximum value of about $1.5\times 10^{-4}$. It indicates that the nonlinearity of the algebraic Reynolds stress model is relatively weak in the ONERA M6 case.

\begin{figure}[!htb]
    \centering
    \includegraphics[width=1.0\textwidth]{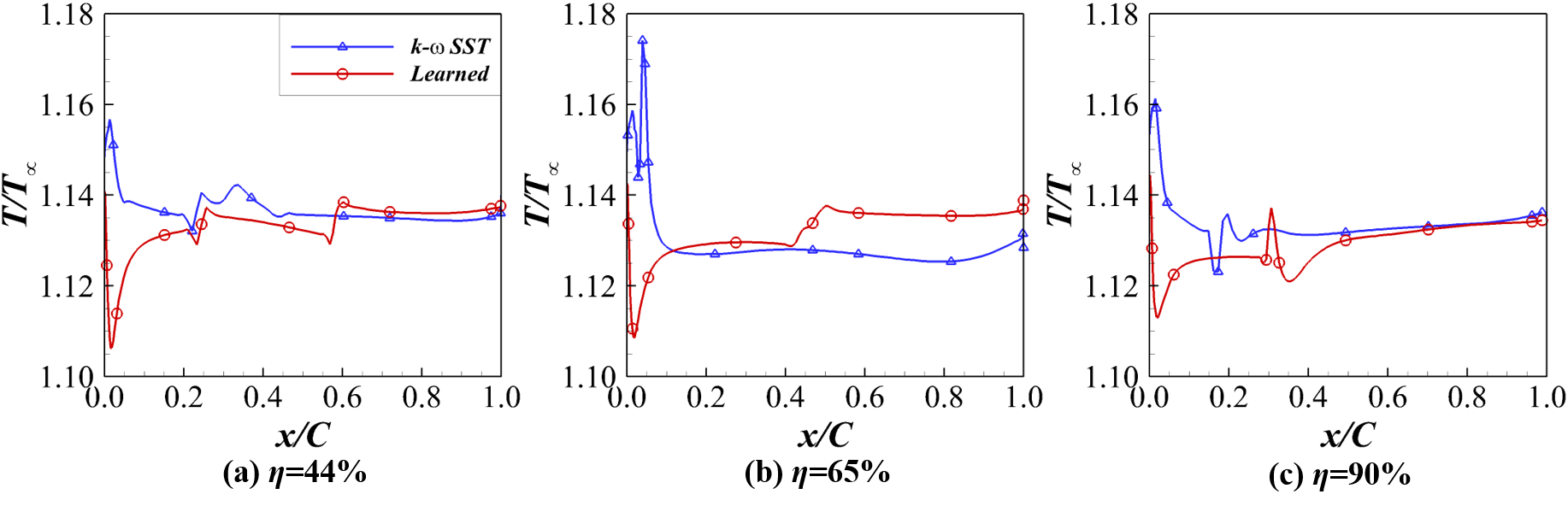}
    \caption{Distribution of temperature on the upper surface of the ONERA M6 wing.}
    \label{fig:M6Temp}
\end{figure}

    Figure~\ref{fig:M6Temp} presents the predicted temperature distribution on the upper surface of the wing with the learned model and the baseline model.
    The temperature distribution on the surface is significantly affected by the location of the shock wave. 
    One can see that the wall temperature rises across the shock wave region, due to the strong compressibility effects. 
    At the section of $\eta=90\%$ in Fig.~\ref{fig:M6Temp} (c), the wall temperature is significantly reduced behind the shock wave.
    That is caused by the momentum transport weakening and further heat production decreasing at the shock-induced separation region.
    Also, it is noticeable that the learned model leads to more severe temperature reductions compared to the baseline model.
    That is likely because the learned model suppresses the turbulent heat transfer by reducing the magnitude of $g^{(1)}$ coefficient at the post-shock region as shown in Fig.~\ref{fig:M6Gfunction}(c).

\begin{figure}[!htb]
    \centering
    \includegraphics[width=1.0\textwidth]{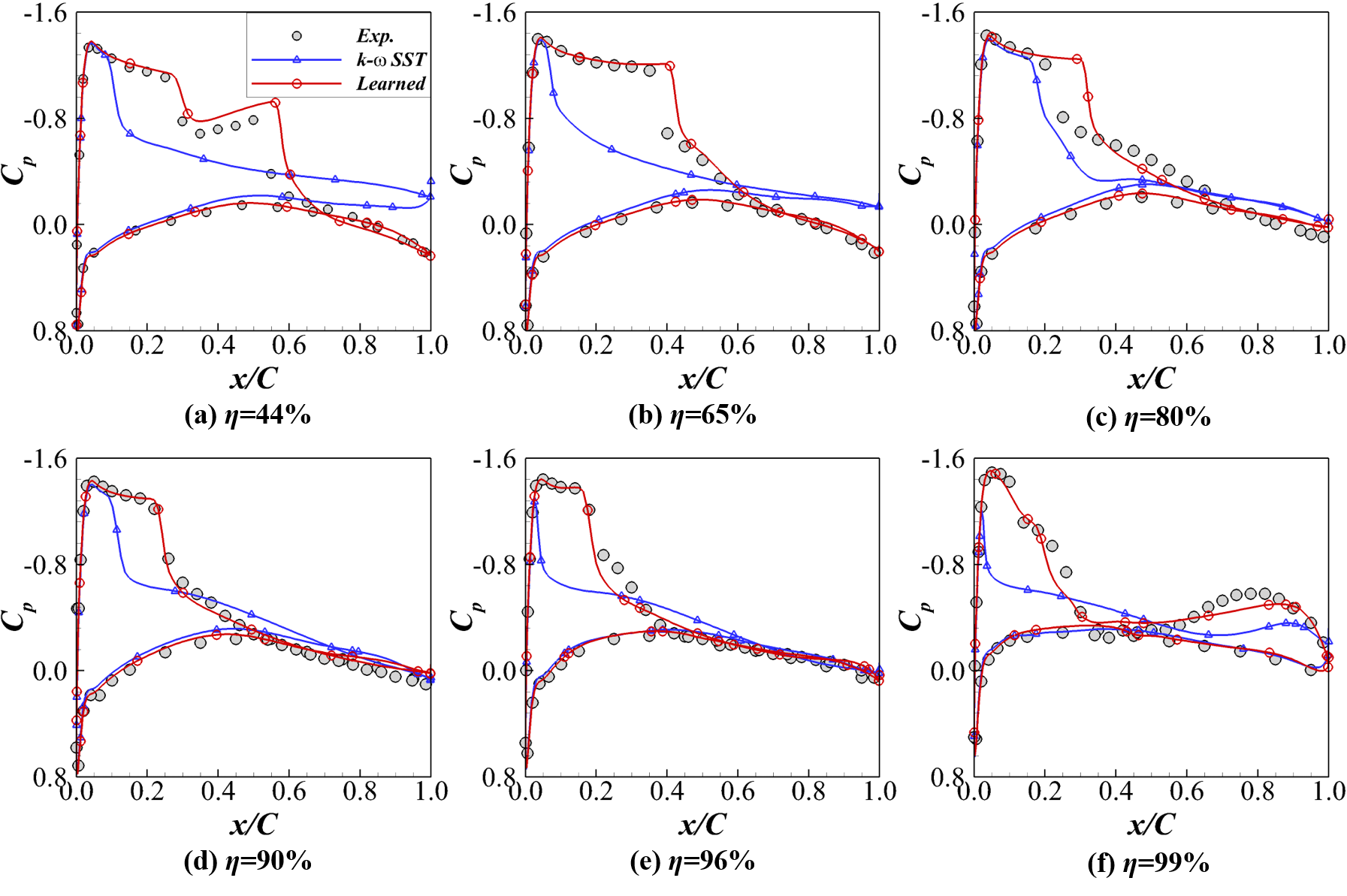}
    \caption{ Results of generalization tests in surface pressure coefficients with comparison among the baseline model, the learned model, and experimental data, for the ONERA M6 case.}
    \label{fig:M6CpFrst}
\end{figure}

    The learned model can be generalized well to different flow conditions for the ONERA M6 case.
    Specifically, the learned model is tested for the ONERA M6 wings at the flow condition of $Ma=0.84$, $\alpha = 6.06^\circ$, and Reynolds number based on the mean aerodynamic chord $Re_{MAC} = 1.17 \times 10^7$.
    This flow has a wider flow separation zone than the training case in spite of slight variations in the angle of attack.
    The predicted pressure coefficients with the baseline model and the learned model are shown in Fig.\ref{fig:M6CpFrst} with comparison to the experimental measurements.
    It shows that the learned model achieves more accurate predictions than the baseline model.
    In addition, the learned model is also tested for the ONERA M6 wings at the flow condition of $Ma=0.84$, $\alpha = 3.06^\circ$, and $Re_{MAC} = 1.17 \times 10^7$. 
    This test case is noticeably different from the training case since the shock wave on the suction side is relatively weak and the flow is almost attached without separation. 
    For such attached flows, both the baseline model and the learned model can obtain good agreement with the experimental data, and hence the plots are omitted for brevity.
    The error between the prediction and experiment for the training and test cases is summarized in Table~\ref{tab:ErrorEstmtM6}, quantitatively showing the better predictive performance of the learned model than the baseline model in both training and test cases. 
    Here the error is computed based on the results at the spanwise locations $\eta=44\%, 65\%, 80\%, 90\%, 96\%, 99\%$.

\begin{table}[!hbt]
    \caption{\label{tab:ErrorEstmtM6} 
    Summary of the prediction error in wall pressure coefficients for the ONERA M6 case.}
    \centering
    \begin{tabular}{ccc} 
    \hline
    \hline
      Cases             & Turbulence models  & Error estimation of $C_p$    \\  
    \hline
    \multirow{2}{*}{Training case}     & $k$--$\omega$ SST              & $35.767 \%$    \\ % 3.432827 
                                       & learned model               & $12.378 \%$    \\ % 1.188094 
    \multirow{2}{*}{Test case (1)}     & $k$--$\omega$ SST              & $36.807 \%$    \\  % 3.709330 
                                       & learned model               & $15.300 \%$    \\  % 1.541848 
    \multirow{2}{*}{Test case (2)}  & $k$--$\omega$ SST    & $12.356 \%$    \\   
                                     & learned model    & $11.645 \%$  \\  % 1.30316 
    \hline
    \hline
    \end{tabular}
\end{table}

\section{Conclusion}
\label{sec:4-Conclusion}
In this work, we demonstrate the application of the ensemble Kalman method in learning turbulence models for transonic flows from sparse experimental data.
A neural network-based turbulence model for transonic flows is developed by taking into account fluid compressibility.
The normalization strategies of input features, including the min-max normalization and local normalization, are investigated with emphasis on the effects of shock waves on flow features.
Moreover, turbulent heat fluxes are modified with the tensor basis neural networks based on the Boussinesq assumption.
The proposed neural network framework  is tested in two external flow cases, i.e., 2D transonic flows around the RAE2822 airfoil and 3D transonic flows around the ONERA M6 wings.
The experimental measurements of velocity and wall pressure are used to train the neural network-based model, respectively, highlighting the flexibility of the ensemble Kalman method in the use of training data.
Both training and test cases show that the learned model can achieve predictions in good agreement with the training data, which demonstrates the capability of the ensemble-based turbulence modeling framework for transonic flows.

Future works will be conducted to evaluate the machine-learned turbulence models by comparison to classical nonlinear eddy viscosity models~\cite{craft1996dev,rubinstein1990nonlinear} in various applications.
Moreover, the nonlinear eddy viscosity model is still under the weak equilibrium assumption.
The present ensemble method can be extended in the Reynolds stress transport model framework to construct the neural network-based model for the pressure-strain-rate correlation.
Besides, the heat flux closure is based on the gradient diffusion model in this work.
The generalized gradient diffusion model and additional invariants associated with the temperature gradient need to be considered for predicting the heat flux in future investigations.

\appendix

\counterwithin{figure}{section}
\counterwithin{table}{section} %
\counterwithin{equation}{section} %
\renewcommand\thefigure{\Alph{section}.\arabic{figure}}
\renewcommand\thetable{\Alph{section}.\arabic{table}}
\renewcommand\theequation{\Alph{section}.\arabic{equation}}

\section{Baseline $k$--${\omega}$ SST turbulence model}
   \label{AppendixSST}

    The $k$--$\omega$ Shear Stress Transport (SST) model~\cite{menter1994two} is used to evaluate the turbulent kinetic energy $k$ and specific turbulence dissipation rate $\omega$. 
    The corresponding transport equations are
    
\begin{equation}  
    \frac{{\partial \rho k}}{{\partial t}} + \frac{{\partial \left( {\rho {u_j}k} \right)}}{{\partial {x_j}}} = P -  \beta^* \rho k\omega  + \frac{\partial }{{\partial {x_j}}}\left[ {\left( {\mu  + {\sigma _k}{\mu _t}} \right)\frac{{\partial k}}{{\partial {x_j}}}} \right],
    \label{Eq:k-equation}
\end{equation}
\begin{eqnarray}
    \label{eq:w-equation}
    \frac{{\partial \rho \omega }}{{\partial t}} + \frac{{\partial \left( {\rho {u_j}\omega } \right)}}{{\partial {x_j}}} = & \frac{{\gamma \rho }}{{{\mu _t}}}P - \beta \rho {\omega ^2} + \frac{\partial }{{\partial {x_j}}}\left[ {\left( {\mu  + {\sigma _\omega }{\mu _t}} \right)\frac{{\partial \omega }}{{\partial {x_j}}}} \right] \nonumber\\
    & + 2\rho \left( {1 - {F_1}} \right)\frac{{{\sigma _{\omega 2}}}}{\omega }\frac{{\partial k}}{{\partial {x_j}}}\frac{{\partial \omega }}{{\partial {x_j}}} \text{.}
\end{eqnarray}

    The production of turbulent kinetic energy is stress-based and defined as
\begin{equation}  
    P = {\tau _{ij}}\frac{{\partial {u_i}}}{{\partial {x_j}}} \text{,}
    \label{Eq:k-production}
\end{equation}
    where the Reynolds stress tensor $\tau _{ij}$ is obtain by the proposed neural-network-based turbulence model. The blending function $F_1$ is given by 
\begin{equation}  
    \left\{ 
    \begin{gathered}
    \begin{array}{*{20}{c}}
        {{F_1} = \tanh \left( {{\Gamma ^4}} \right),}&{\Gamma  = \min \left[ {\max \left( {{\Gamma _1},{\Gamma _3}} \right),{\Gamma _2}} \right]} 
    \end{array} \hfill \\
    \begin{array}{*{20}{c}}
        {{\Gamma _1} = \frac{{500\mu }}{{\rho \omega {d^2}}}}&{{\Gamma _2} = \frac{{4{\sigma _{\omega 2}}\rho k}}{{C{D_{k\omega }}{d^2}}}}&{{\Gamma _3} = \frac{{\sqrt k }}{{{\beta ^*}\omega d}}} 
    \end{array} \hfill \\
    C{D_{k\omega }} = \max \left( {2\rho \frac{{{\sigma _{\omega 2}}}}{\omega }\frac{{\partial k}}{{\partial {x_j}}}\frac{{\partial \omega }}{{\partial {x_j}}},{{10}^{ - 20}}} \right) \hfill  
    \end{gathered}  \right. \text{.}
    \label{Eq:BlendingF}
\end{equation}  
    Here, $d$ is the shortest distance from the wall. Each of the constants in Eqs.~\eqref{Eq:k-equation} and ~\eqref{eq:w-equation} is blended via
\begin{equation}  
    \varphi  = {F_1}{\varphi _1} + \left( {1 - {F_1}} \right){\varphi _2} \text{.}
    \label{Eq:BlendingEquatn}
\end{equation}  
    The constants are
\begin{equation}  
    \begin{gathered}
    \begin{array}{*{20}{c}}
        {{\sigma _{k1}} = 0.85}&{{\sigma _{k2}} = 1.0,}&{{\sigma _{\omega 1}} = 0.5,}&{{\sigma _{\omega 2}} = 0.856} \text{,} 
    \end{array} \\ 
    \begin{array}{*{20}{c}}
        {{\beta _1} = 0.075}&{{\beta _2} = 0.0828,}&{{\beta ^*} = 0.09} \text{,}   
    \end{array} \\ 
    \begin{array}{*{20}{c}}
        {\kappa  = 0.41}&{{\gamma _1} = \frac{{{\beta _1}}}{{{\beta ^*}}} - \frac{{{\sigma _{\omega 1}}{\kappa ^2}}}{{\sqrt {{\beta ^*}} }}}&{{\gamma _2} = \frac{{{\beta _2}}}{{{\beta ^*}}} - \frac{{{\sigma _{\omega 2}}{\kappa ^2}}}{{\sqrt {{\beta ^*}} }}} \text{.}
    \end{array}
    \end{gathered}
\end{equation} 
    More details can be found in ref.~\cite{menter1994two}.

\section{Sensitivity analysis of mesh grid and  network architecture}
    \label{sec:Psnesitivity}

\begin{figure}[!htb]
    \centering
    \subfloat[$C_p$-distribution]{
    \includegraphics[width=0.45\textwidth]{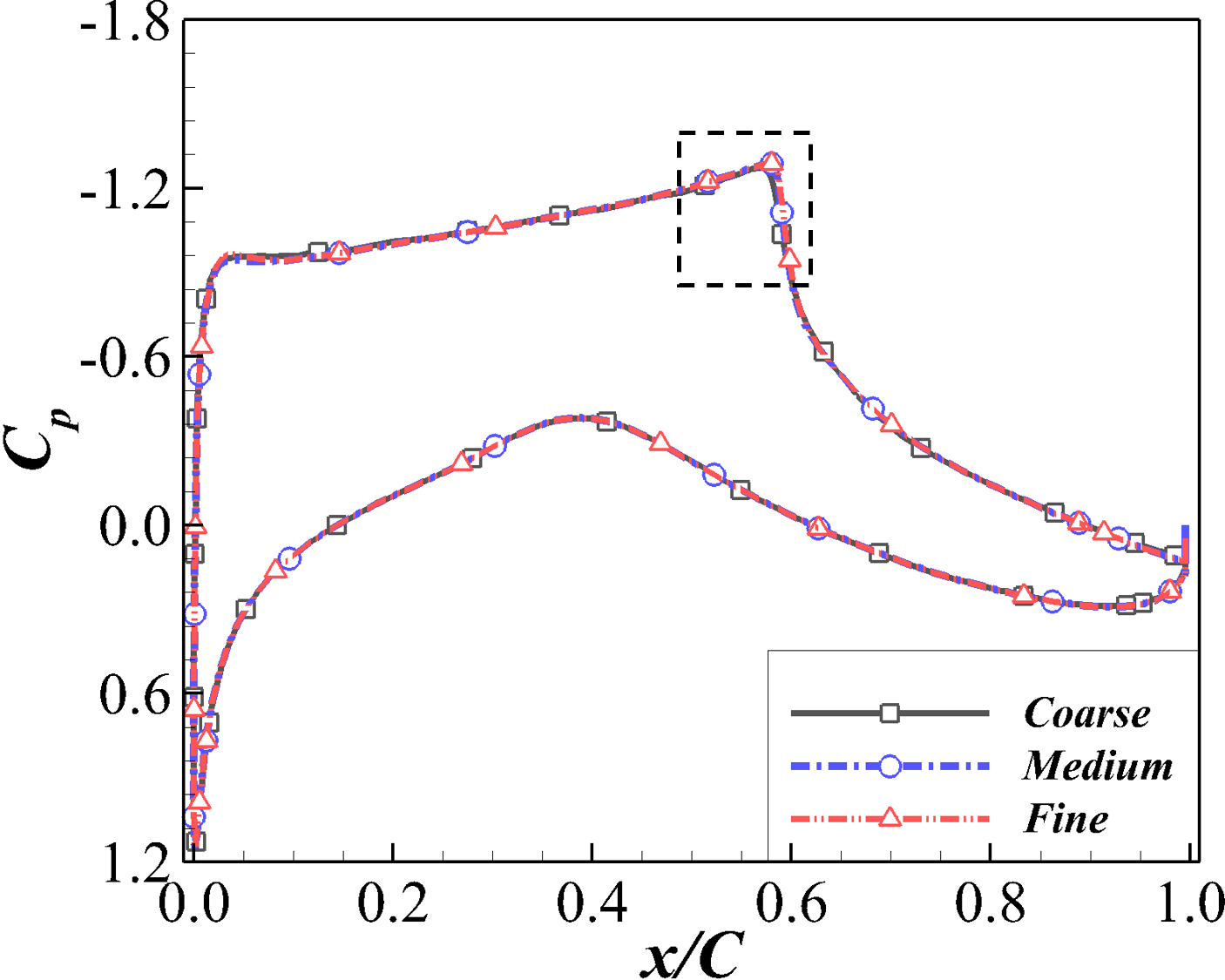} }
    \hspace{3mm}
    \subfloat[Zoomed in view]{
    \includegraphics[width=0.45\textwidth]{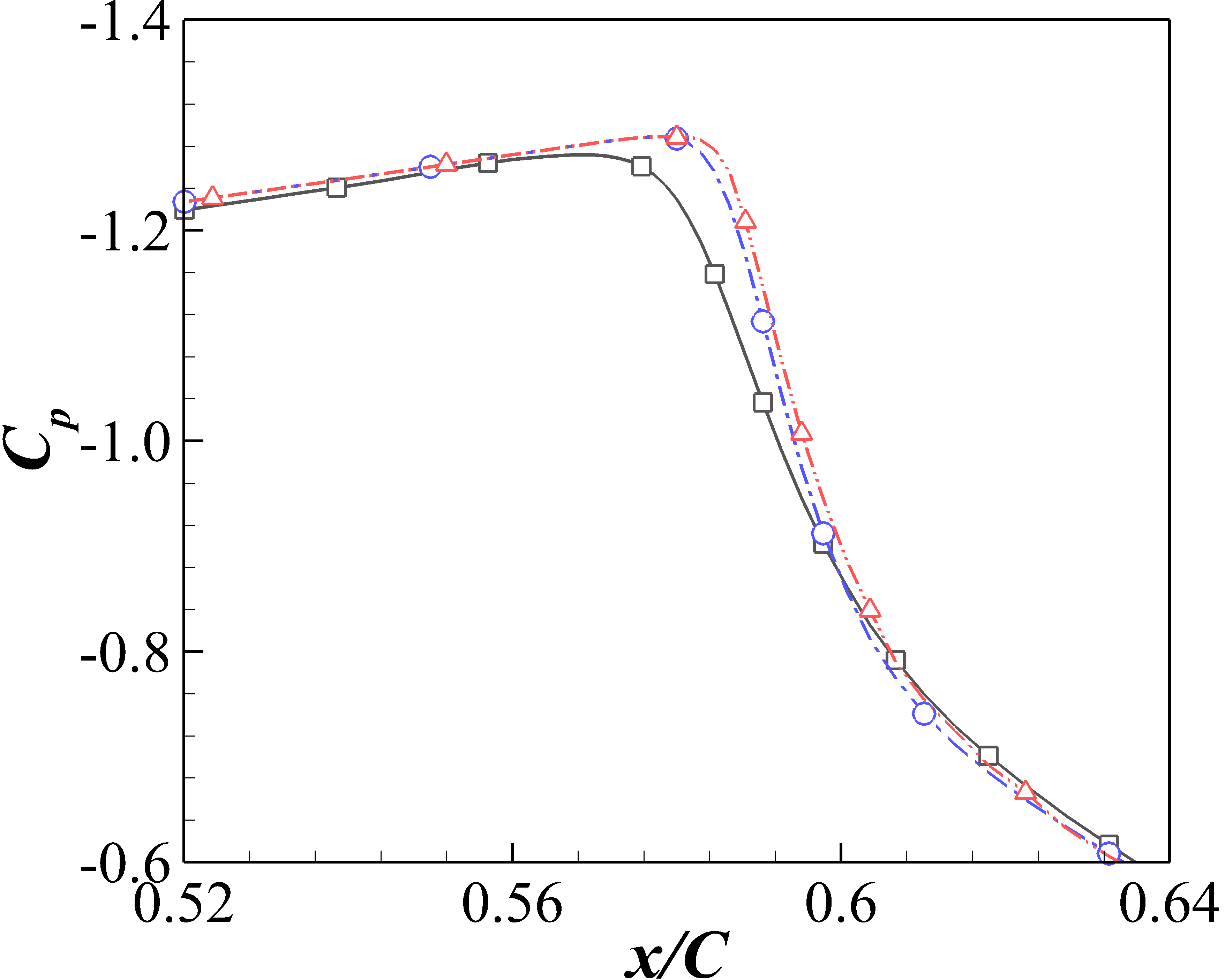} }
    \caption{Comparison of predicted wall pressure distribution with various mesh sizes for the RAE 2822 cases.} 
    \label{figA1:RAE2822MeshStudy}
\end{figure}

    %1.mesh
    In the RAE2822 cases, the mesh sensitivity is investigated for the training case at the flow condition of $Ma=0.75$, $\alpha={2.8^\circ}$, and $Re_C = 6.2 \times 10^6$. 
    Three typical meshes with different resolutions are generated for comparison, and the first layer spacing of all meshes in the normal direction corresponds to ${y^+} \approx 0.8$.
    The cell numbers of these meshes are summarized in Table~\ref{tab:DetailMeshFor2822},
    and all the grids are generated with the same topology. 
    Figure~\ref{figA1:RAE2822MeshStudy} compares the predicted wall pressure distribution~$C_p$ with various mesh sizes. 
    The wall pressure distribution is affected by the mesh resolution mainly near the shock location, which exhibits that the medium and fine meshes provide very close results. 
    Hence, the medium-density mesh is adopted for the RAE2822 case.

\begin{table}[!htb]
    \caption{\label{tab:DetailMeshFor2822} Summary of mesh grids for mesh sensitivity study of the RAE 2822 case.}
    \centering
    \begin{tabular}{cccc}
    \hline
    \hline
    Meshes           &   Coarse   &   Medium   &  Fine   \\
    \hline
    Number of cells  &  $32,047$  &  $45,394$  & $104,122$ \\
    \hline
    \hline
    \end{tabular}
\end{table}

\begin{figure}[!htb]
    \centering
    \includegraphics[width=0.8\textwidth]{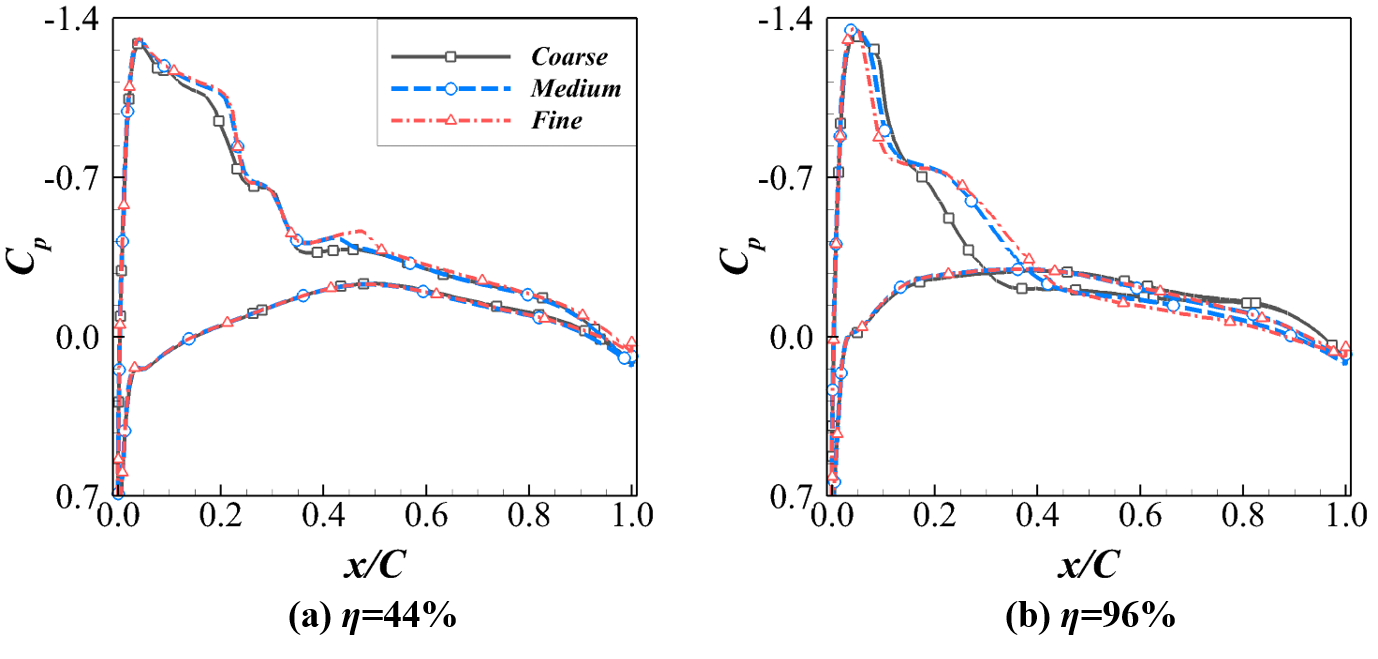}
    \caption{Comparison of predicted wall pressure distribution with various mesh sizes for the ONERA M6 case.} 
    \label{figA1:M6MeshStudy}
\end{figure}

    We also perform the mesh sensitivity study for the ONERA M6 cases.
    Three meshes are generated with the same topology, and the cell numbers of each mesh are summarized in Table~\ref{tab:DetailMeshForM6}.
    All generated meshes are simulated with the training case, i.e., at the flow condition of $Ma=0.84$, $\alpha={5.06^\circ}$ and ${\rm{Re = 1}}{\rm{.17}} \times {10^7}$. 
    Figure~\ref{figA1:M6MeshStudy} shows the comparisons of simulated wall pressure distribution with various mesh sizes.
    The variation of mesh density affects the wall pressure distribution on the suction side, especially at the section of $\eta  = 96\% $.
    The differences between the medium and the fine mesh are small in the predicted wall pressure.
    Therefore, the medium-density mesh is used for the ONERA M6 case.

\begin{table}[!htb]
    \caption{\label{tab:DetailMeshForM6} Summary of mesh grids for mesh sensitivity study of ONERA M6 cases.}
    \centering
    \begin{tabular}{cccc}
    \hline
    \hline
    Meshes           &    Coarse     &    Medium     &   Fine       \\
    \hline
    Number of cells  &  $2,735,233$  &  $3,627,926$  & $6,563,867$  \\
    \hline
    \hline
    \end{tabular}
\end{table}

    %2.NN
    The sensitivity to the architecture of neural networks is investigated for both the RAE 2822 case and the ONERA M6 case.
    Four neural network architectures are tested: (1) two hidden layers with 5 neurons per layer;
    (2) two hidden layers with 10 neurons per layer;
    (3) four hidden layers with 20 neurons per layer;
    (4) eight hidden layers with 10 neurons per layer.
    The results of prediction errors for the RAE2822 case and the ONERA M6 case are summarized in Tables~\ref{tab:NnNhFor2822} and~\ref{tab:NnNhForM6}, respectively.
    % Generally, the neural network architecture has no significant influence on the training accuracy.
    For the RAE2822 case, the neural network with two hidden layers and $5$ neurons per layer provides the best predictions in the velocity compared to the other neural network architectures.
    Moreover, this neural network predicts the pressure coefficient with the error of $7.701\%$, which  has only subtle differences from other cases.
    As for the ONERA M6 case, the neural network with two hidden layers and 10 neurons per layer provides the best estimation of the wall pressure coefficient~$C_p$, compared to the other architectures.
    Therefore, the neural network with two hidden layers and 5 neurons per layer is used for the RAE2822 case, and the neural network with two hidden layers and 10 neurons per layer is adopted for the ONERA M6 case.
    
\begin{table}[!htb]
    \caption{\label{tab:NnNhFor2822} Parameter sensitivity analysis of neural networks for the RAE 2822 cases.}
    \centering
    \begin{tabular}{ccccc}
    \hline
    \hline
    \multirow{2}{*}{Hidden layers}  &  \multirow{2}{*}{Neurons per layer}  &  \multirow{2}{*}{Trainable parameters}  &  \multicolumn{2}{c}{Error estimation}  \\  
    &              &                                & Velocity  &  $C_p$   \\
    \hline
    $2$ & $5$  & $68$   & $7.868\%$ & $7.701\%$ \\
    % \normalsize{\textcircled{\scriptsize{2}}} & 
    $2$ & $10$ & $183$  & $8.262\%$ & $7.656\%$ \\
    % \normalsize{\textcircled{\scriptsize{3}}} & 
    $4$ & $20$ & $1403$ & $8.463\%$ & $7.902\%$ \\
    % \normalsize{\textcircled{\scriptsize{4}}} & 
    $8$ & $10$ & $843$  & $8.069\%$ & $7.864\%$ \\
    \hline
    \hline
    \end{tabular}
\end{table}

\begin{table}[!htb]
    \caption{\label{tab:NnNhForM6} Parameter sensitivity analysis of neural networks for the ONERA M6 cases.}
    \centering
    \begin{tabular}{cccc}
    \hline
    \hline
    Hidden layers  &  Neurons per layer  &  Trainable parameters &  Error estimation of $C_p$  \\
    \hline
    $2$  & $5$   &  $68$  & $12.823\%$ \\
    $2$  & $10$  & $183$  & $12.378\%$ \\
    $4$  & $20$  & $1403$ & $12.460\%$ \\ 
    $8$  & $10$  & $843$  & $12.668\%$ \\ 
    \hline
    \hline
    \end{tabular}
\end{table}

\section*{Acknowledgment}
This work is supported by the NSFC Basic Science Center Program for ``Multiscale Problems in Nonlinear Mechanics'' (No. 11988102), the National Natural Science Foundation of China (Nos. 12102439 and 12102435), and the China Postdoctoral Science Foundation (Nos. 2021M703290 and 2021M690154).
The authors would like to thank the reviewers for their constructive and valuable comments, which helped improve the quality and clarity of this manuscript.

\end{document}